\documentstyle[12pt,draft,aas_macros]{nature}

\font\large=cmbx10 scaled \magstep3

\newcommand{\equ}[1]{eq.~(\ref{eq:#1})}

\newcommand{\fig}[1]{Fig.~\ref{fig:#1}}
\newcommand{\Fig}[1]{Fig.~\ref{fig:#1}}
\newcommand{\be}{\begin{equation}}
\newcommand{\ee}{\end{equation}}

\newcommand{\msun}{M_\odot}
\newcommand{\ifm}[1]{\relax\ifmmode#1\else$\mathsurround=0pt #1$\fi}
\newcommand{\kms}{\ifmmode\,{\rm km}\,{\rm s}^{-1}\else km$\,$s$^{-1}$\fi}
\newcommand{\kpc}{\ifmmode\,{\rm kpc}\else kpc\fi}
\newcommand{\Mpc}{\ifmmode\,{\rm Mpc}\else kpc\fi}

\newcommand{\ltsima}{$\; \buildrel < \over \sim \;$}
\newcommand{\lsim}{\lower.5ex\hbox{\ltsima}}
\newcommand{\gtsima}{$\; \buildrel > \over \sim \;$}
\newcommand{\gsim}{\lower.5ex\hbox{\gtsima}}
\newcommand{\prop}{\propto}

\def\Re{R_{\rm eff}}
\def\Rv{R_{\rm vir}}

\def\rp{r_{\rm p}}
\def\sr{\sigma_r}
\def\st{\sigma_\theta}
\def\sp{\sigma_{\rm p}}
\def\vp{v_{\rm p}}

\def\no{\noindent}

\def\mno{\medskip\noindent}

\newcommand{\ad}[1]{}

\begin{document}

\title{Lost \& found dark matter in elliptical galaxies} 

\author{A. Dekel$^{1,2,3}$, F. Stoehr$^2$, G.A. Mamon$^2$, T.J. Cox$^4$,
        G.S. Novak$^5$, 
        \& J.R. Primack$^3$
\institute{$^1$Racah Institute of Physics, 
               The Hebrew University, Jerusalem 91904, Israel\\
$^2$Institut d'Astrophysique and Observatoire de Paris, 
98bis Boulevard Arago, Paris 75014, France\\
$^3$Department of Physics, University of California, Santa Cruz CA 95064, USA\\
$^4$Center for Astrophysics, Harvard University, 60 Garden Street, Cambridge MA
02138, USA\\
$^5$UCO/Lick Observatories, University of California, Santa Cruz CA 95064, USA
}
}

\dates{}{}
\mainauthor{Dekel et al.}
\headertitle{Dark matter in elliptical galaxies}

%%%%%%%%%%%%%%%%%%%%%%%%%%%%%%%%%%%%%%
%1
\summary{There is strong evidence that the mass in the Universe 
is dominated by dark matter, which exerts gravitational attraction but 
whose exact nature is unknown.
In particular, all 
galaxies are believed to be embedded in massive haloes of dark 
matter\cite{sofue01,wr78}. 
This view has recently been challenged by surprisingly low random 
stellar velocities in the outskirts of ordinary elliptical galaxies, 
which were interpreted as indicating a lack of dark 
matter\cite{mendez01,romanowsky03}. 
Here we show that the low velocities 
are in fact compatible with galaxy formation in dark-matter haloes.  
Using numerical simulations of disc-galaxy mergers\cite{cox04,cox05}, 
we find that the stellar orbits in the outer regions of the resulting 
ellipticals are very elongated. These stars were torn by tidal forces from 
their original galaxies during the first close passage and put 
on outgoing trajectories. The elongated orbits, combined with the  
steeply falling density profile of the observed tracers, explain 
the observed low velocities even in the presence of large 
amounts of dark matter. Projection effects when viewing 
a triaxial elliptical can lead to even lower observed velocities along 
certain lines of sight.
}

\maketitle

%%%%%%%%%%%%%%%%%%%%%%%%%%%%%%%%%%%%%%
\noindent{\bf Introduction}

%2
\noindent
% CDM paradigm. Evidence in discs. Theory reuires DM
The common spiral galaxies are known to reside in extended 
dark-matter (DM) haloes.  The rotational speeds of their gas 
discs do not decline outside the visible body\cite{sofue01}, 
unlike the expectation from Keplerian 
circular %D 
velocities at a radius 
$r$ about a mass $M$, $V^2=G M/r$.  Thus, the DM mass 
within $r$ is growing roughly as $M(r)\propto r$ and it dominates 
the gravitational potential beyond a certain radius.  An extrapolation based 
on the typical halo density profile\cite{nfw97} found in simulations of 
the standard $\Lambda$CDM cosmology predicts an outer ``virial" radius $\Rv$ 
that is 50-100 times larger than the characteristic stellar radius, 
enclosing 10-20 times more DM than luminous matter,
now also indicated observationally.\cite{prada03}. %D   
The conventional wisdom is that the potential wells created by the DM 
are crucial for seeding the formation of galaxies\cite{wr78,blum84,ds86}.

%3
% Partial evidence in E's
The standard hypothesis is that ellipticals originate from mergers 
of discs\cite{fall79} and should therefore be embedded in similar DM 
haloes.  There is evidence for DM in giant ellipticals, from 
X-rays\cite{mathews03} and gravitational lensing\cite{keeton01} 
and in nearby dwarf galaxies from stellar kinematics \cite{mateo98}. %D 
However, ordinary ellipticals lack obvious velocity tracers at the 
large projected radii $\rp$ where the DM is expected to be important.  
This is typically beyond $\Re$,\cite{mamon05b} the ``effective" radius 
encompassing half the 
projected %D 
light, while measurements of the projected 
velocity dispersion $\sp$ of the stellar light are limited to $\rp < 2\Re$. 

%4
% PN -> no haloes in Es. A crisis for CDM?  MOND?
The strong [O{\small III}] emission line at 5007${\rm\AA}$ from Planetary 
Nebulae (PN) --- hot shells of gas expelled from dying stars of 
$(1-3)\msun$ --- provide a unique tool for extracting $\sp(\rp)$ beyond $\Re$. 
Romanowsky~et al.\cite{romanowsky03} measured $\sp$ from 
$\sim$100 bright PNs in each of three normal ellipticals, 
NGC 821, 3379 and 4494, 
adding to 531 PNs in NGC 4697 by Mendez et al.\cite{mendez01}. 
They find that $\sp$ typically drops by a factor $\simeq 1.6$ between 
$\rp = \Re$ and $3\Re$.
%The $\sp$ of PNs in the normal ellipticals
%NGC 821, 3379 and 4494\cite{romanowsky03} and in NGC 4697\cite{mendez01} 
%were found to typically drop by a factor $\simeq 1.6$ between 
%$\rp = \Re$ and $3\Re$.
The kinematic modelling by the observers\cite{romanowsky03} yielded low
mass-to-light ratios, e.g. $M/L \simeq 7$ at $5\Re$ for NGC 3379, 
consistent with a ``naked" stellar population.  They interpreted this as 
``little if any DM in these galaxies' haloes",
in conflict with the conventional picture. %D
While noticing that increasing velocity anisotropies could 
in principle produce declining $\sp$, they ruled out such 
``pathological" orbit structure.  
Similar conclusions were obtained later from other 
ellipticals\cite{napolitano05}.  
The apparent challenge to theory has already triggered radical 
explanations\cite{milgrom03}.  However, the 
earlier analysis\cite{romanowsky03} 
might have missed alternative solutions because it was limited to 
specific density-profile shapes in stationary spherical systems, the halo 
PNs were identified with the central stellar population, and their 
maximum-likelihood method may suffer from an incomplete orbit library 
or questionable convergence~properties\cite{valluri04}.
%D Cretton & Emsellem 2004, MNRAS 347, L31
%D richstone05 astro-ph/0403257 ApJL
% 190 w  -13

%%%%%%%%%%%%%%%%%%%%%%%%%%%%%%%%%%%%%%%%
\bigskip
\noindent{\bf Anisotropic velocity dispersion}

%5
\noindent
% why high beta gives low sp: intuitive
For given density profiles, a lower $\sp$ can result from 
more radial velocities. The dynamics implies a lower 3D velocity
dispersion $\sigma$ because the pressure needed for balancing gravity is 
provided by a radial $\sr$ that corresponds to a lower 
$\sigma$.  The projection introduces a further decrease in $\sp$.  This 
can be illustrated by toy models made purely of circular 
or radial orbits, with the same constant speed and random 
orientations.  If the stellar-density profile is steep enough, $\sp$ 
is dominated by the tangential contribution near the equatorial plane 
perpendicular to the line of sight, which is high for 
circular orbits and low for radial orbits.

%6
% more math
%The collisionless Boltzmann equation implies that 
The 3D profiles of {\it any} component of a spherical gravitating 
system in equilibrium obey the Jeans equation\cite{bt87}
(derived from the collisionless Boltzmann equation), %D
\be
V^2(r) = [\alpha(r)+\gamma(r)-2\beta(r)]\, \sr^2(r) \ ,
\label{eq:jeans}
\ee
a manifestation of 
local hydrostatic balance between the inward pull of gravity (left) 
and the outward push of pressure (right).  Here $V^2(r)=GM(r)/r$ is 
the squared circular velocity.  The stellar density profile $\nu(r)$ enters 
via $\alpha\equiv -d\ln \nu/d\ln r$.  Its velocity dispersion consists of 
radial and tangential 
components, $\sigma^2=\sr^2+2\st^2$; we define 
$\gamma\equiv -d\ln \sr^2/d\ln r$.  
The velocity anisotropy is 
$\beta \equiv 1-\st^2/\sr^2$, with $\beta=-\infty,0,1$ for circular, 
isotropic and radial orbits respectively.  
The projection can be performed analytically when $\beta$, $\alpha$ and 
$\gamma$ are constant with $r$ (power-law profiles, 
$V^2=V_0^2 (r/\Re)^{-\gamma}$): 
\be
\sp^2(\rp)= A(\alpha,\gamma)\, 
\left(\frac{(\alpha+\gamma)-(\alpha+\gamma-1)\beta}
{(\alpha+\gamma)-2\beta}\right)
V_0^2\, \left( \frac{\rp}{\Re} \right) ^{-\gamma} 
\label{eq:sp}
\ee
($A$ in Methods).  Note that $\sp$ is a decreasing function 
of $\beta$ and of $\alpha$ (for $\alpha+\gamma>3$ and $\beta>0$)
respectively. %D 
Local fits to the de-projection of the standard de Vaucouleurs\cite{devac48} 
surface-brightness profile of ellipticals give $\alpha \simeq 3.13$-3.37 
at 2-$3\Re$.\cite{mamon05a}
% (or 2.28-2.52 in 2D)  3.2 at 2.26Re   3.26 at 2.5Re
Our fits to $\sp^2(\rp)$ in the observed ellipticals beyond $\Re$ yield 
$\gamma \simeq 0.8\pm0.2$.  
%D With $\alpha+\gamma\simeq 4.1$ we get 
These give 
$A(\alpha,\gamma)\simeq 0.2$ and $\sp$
%D $\sp \simeq 0.45 [(4.2-3.2\beta)/(4.2-2\beta)]^{1/2} V_0 (\rp/\Re)^{-0.45}$,
drops by a factor $\sim 1.5$ between $\beta=0$ and $1$.  
We learn that one could match the low $\sp$ 
at large radii either by a low $V_0$ or by a high $\beta$
there, and that a high $\alpha$ helps.

%7
% model fits
In a more realistic model, we assume a S\'ersic stellar density
profile\cite{sersic68,caon93,mamon05a} %D
%D \cite{caon93} %D need to cut to 30 refs
(= de Vaucouleurs' for S\'ersic index $m=4$), a standard DM 
density profile\cite{nfw97} with a typical concentration 
($\sim 10$)\cite{bullock01_c}, and a virial stellar mass fraction 
$\sim 0.06$.  With $\beta=0$ we recover the 
apparent 5D 
discrepancy, but $m \simeq 4$ and $\beta(r>\Re) \simeq 0.5$ 
(independent of its behaviour well inside $\Re$) 
yield a $\sim 1\sigma$ agreement with the observed $\sp$. 
A good fit is obtained with either $m \simeq 4$ and $\beta \simeq 0.75$, 
or with $m \simeq 2.3$ (for which $\alpha \simeq 3.5$ near $2.5\Re$) and 
$\beta \simeq 0.5$.
The required $\beta$ is higher than the $\beta \leq 0.2$ predicted for DM
particles %D 
in typical haloes at a few $\Re$ ($\ll\Rv$)\cite{mamon05b}, %D 
so the PNs must not trace the DM kinematics.

%%%%%%%%%%%%%%%%%%%%%%%%%%%%%%%%%%%%%%%%%%%
\bigskip
\noindent{\bf Merger simulations}

\begin{figure}
\vskip 11.6cm 
{\includegraphics{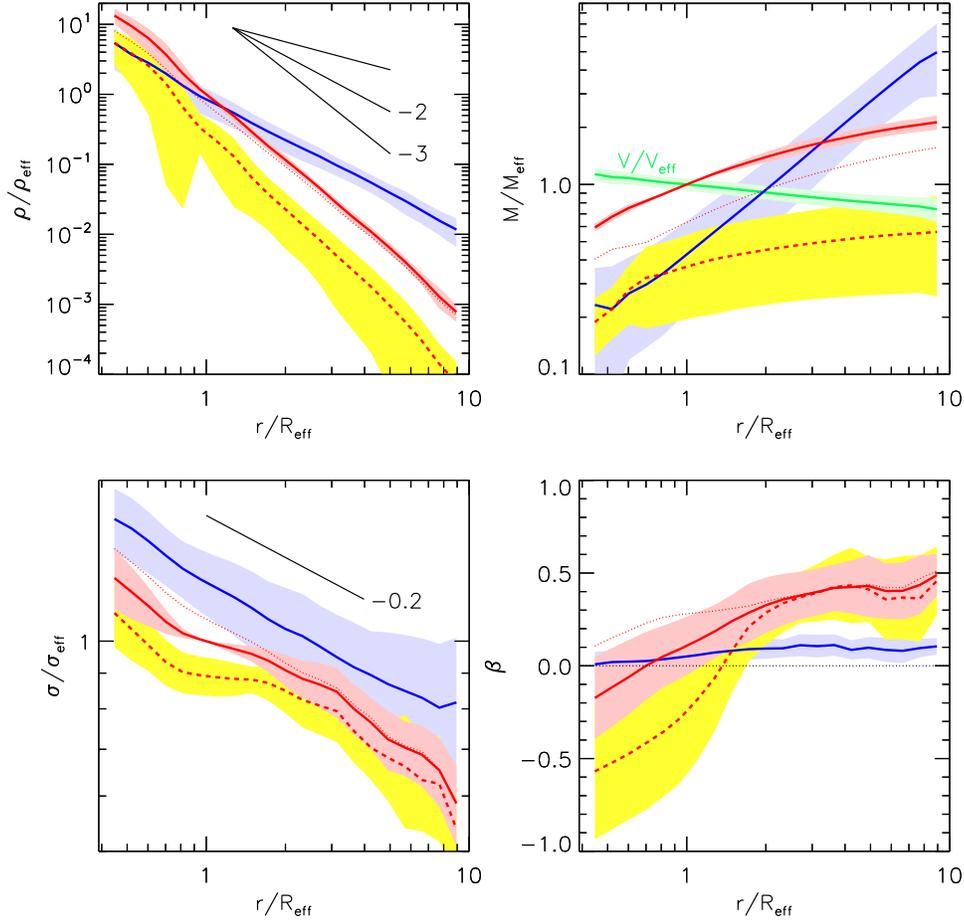}}
\vskip 0.2cm
\caption{Three-dimensional profiles of the simulated merger remnants.
Ten galaxies at two different times after the merger (typically 0.8 and
1.3 Gyr) are stacked together.    
Shown are the profiles for the
dark matter (blue) and the stars (red),
divided into the old ones from the progenitors (dotted) and
those newly formed during the merger (dashed).  The scaling is such
that the curves for the stars (all, solid red) are matched at $\Re$.
The shaded areas mark $1\sigma$ scatter.  The panels refer to density
$\rho$, mass $M$ and circular velocity $V$, velocity dispersion $\sigma$
and anisotropy $\beta$, with ``eff" referring to 
the quantities at   
$\Re$.}
\label{fig:3d}
\end{figure}

%8
\noindent
Assuming that ellipticals form by mergers, and that major mergers of discs 
can reveal generic features of mergers in the $\Lambda$CDM cosmology, 
we appeal to a suite of simulations of such events\cite{cox04,cox05}.
Two spirals are put on a parabolic orbit, 
each consisting of stellar and gaseous discs and a bulge, 
all embedded in a $\Lambda$CDM halo, constructed to match a range
of typical disc galaxies.  The gravitational and hydrodynamical evolution is 
followed using an SPH code\cite{springel01_gadget}, including gas cooling, 
star formation and supernova feedback (Methods).

%9
\Fig{3d} shows the stacked 3D profiles of the merger remnants.
The DM density profile is slightly flatter than 
an isothermal sphere, %D 
$\rho \prop r^{-2}$, similar to simulated $\Lambda$CDM haloes after 
they have responded to gas dissipation\cite{gnedin04}.  
The robust stellar density falls off more steeply, 
$\rho \prop r^{-3.2}$, as in elliptical galaxies %D 
obeying the de Vaucouleurs profile near 2-3$\Re$, 
and with $\Re \simeq 0.015\Rv$.  
For the ``young" stars it is somewhat steeper, $\rho \prop r^{-3.5}$.    
The total-to-stellar mass ratio rises from $\simeq 2$ at $3\Re$ to 
$\simeq 14$ at $\Rv$,
corresponding at $5\Re$ to $M/L \simeq 15$ 
(compared to the earlier\cite{romanowsky03} $M/L \simeq 7$)
(both for stellar $M_*/L=6$). 5D 
The 3D $\sigma$ profiles of the DM and stars have similar slopes 
[as in \equ{sp}], 
falling off roughly as 
$\sigma \prop r^{-0.2}$. 

%10
Our main new finding is the high $\beta$ of the stellar halo 
velocities.  While the DM velocities are almost isotropic ($\beta \sim 0.1$), 
the typical stellar $\beta$ grows from small values at $r<\Re$ 
(sometimes negative, reflecting a small disc that is irrelevant 
to the profiles 
beyond $\Re$)  to $\beta \sim 0.5$ at $r>\Re$. 
In one case $\beta > 0.75$, but in another it remains $<0.2$.
Given $V(r)$ in the Jeans equation, the higher $\alpha$ is 
compensated by a higher $\beta$ and a lower $\sigma$.

%11
The simulations demonstrate that the stellar halo originates from tidal
processes during the first pericentre passage.  Some of the halo stars 
are associated with the two cores and the tidal bridge between them; 
they pass near the center at the coalesence before flying 
outward on radial orbits.  Other halo stars first flow out in extended 
tidal tails and later fall back on radial orbits 
[Supplementary Information, hereafter SI].
Indeed, we find that $\beta$ is correlated with the strength of the
tidal interaction; it is higher 
for more head-on collisions and when the spins are aligned with the orbit.

%%%%%%%%%%%%%%%%%%%%%%%%%%%%%%%%%%%%%
\bigskip
\noindent{\bf Projected profiles}

\begin{figure}
\vskip 15.0cm
{\includegraphics{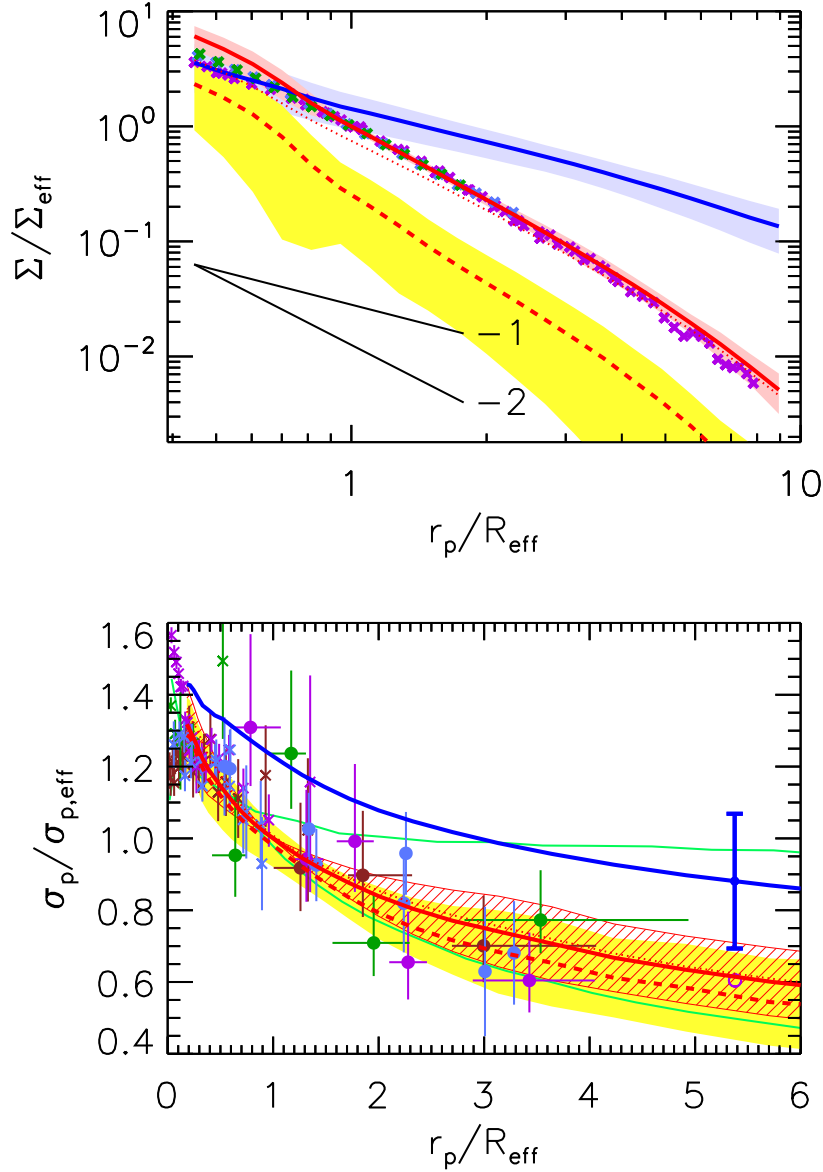}}
\vskip 0.2cm
\caption{
Projected profiles: simulated galaxies versus observations.  Top:
surface density.  Bottom: velocity dispersion.  The merger remnants are viewed
from three orthogonal directions and the 60 profiles are stacked such that the
stellar curves (``all") match at $\Re$.
Colors and line types are as in Fig.~1.
The $<3$ Gyr ``young" stars may mimic the observed PNs.
The $1\sigma$ scatter 
is marked by shaded or hashed areas or a thick bar.
The galaxies are
marked green (821), violet (3379), brown (4494) and blue (4697)
with $1\sigma$ errors;
PNs (circles) and stars (crosses).
The surface densities shown for 3 galaxies almost coincide
with the simulated profile.  Green lines refer to 
the earlier models\cite{romanowsky03} with (upper) and without (lower) DM.
}
\label{fig:2d}
\end{figure}

%12
\noindent
The systems are ``observed" from three orthogonal directions and stacked
together, providing a robust average 
profile and the scatter about it.  The data are scaled 
similarly (Methods).  \Fig{2d} shows the simulated surface-density profile 
and those of NGC 821\cite{goudfrooij94},
3379\cite{devac79} and 4697\cite{peletier90}.  
They are 
all fit by $\Sigma \prop \rp^{-2.3}$ in 
the range 
1-$5\Re$ (as in de Vaucouleurs profile at $\sim 2\Re$).
The simulated projected axial ratios near $\Re$ range 
from 1:1 to 1:2, and the ellipticity is supported by an 
anisotropic, triaxial velocity dispersion rather than by rotation,
similar to ellipticals [SI].  
In addition, 
The distribution of global properties of the 
simulated 
remnants, 
such as luminosity, radius and velocity dispersion, is consistent with 
the ``fundamental plane" of ellipticals [SI].  Thus, the merger remnants 
seem to resemble typical ellipticals near $\sim \Re$ in every relevant respect.

% fig 2
 
%13 
The velocity dispersions in \Fig{2d} demonstrate our punchline.  While the
DM $\sp$ indeed lies above the outer observed points, the 
stellar $\sp$, $\sim 30\%$ lower, provides a good fit everywhere in 
the range 0.5-$4\Re$.  
The slope of the simulated 
$\sp^2(\rp)$ in the range 0.5-$6\Re$ 
is $\gamma=0.53 \pm 0.16$ for ``all" and $\gamma=0.61 \pm 0.22$ 
for the ``young" stars, both consistent with the  
$\gamma=0.59 \pm 0.13$ observed for PNs.
% $\gamma=0.59 \pm 0.08 \pm 0.10$ 
%(error of the mean and scatter) for the 4 galaxies.
No fiddling with model parameters is involved 
--- simply stacking a sample of merger remnants
as simulated.

%14 
% robust to dissipation, major-minor
The $\sp$ of ``young" stars is lower by $\sim 9\%$ at $3\Re$
(due to their larger $\alpha$).  Stellar theory indicates that these 
objects, $<3$ Gyr old, may represent the observed PNs. Emerging from 
1.4-2.5$\msun$ stars\cite{marigo04}, the PNs are expected to be much 
more luminous than those of the older, less massive stars, 
which fall below the detection limits\cite{mendez01,romanowsky03} %D2 
(Methods).  However, the radial orbits 
are a generic result independent of the degree of gas 
dissipation during the merger: our mergers with initial gas-to-baryon 
ratio ranging from 0 to 70\% show negligible differences in 
$\beta$ beyond $\Re$.  Dissipation results in a more centrally concentrated 
stellar distribution, associated with $\lsim 10\%$ reduction in
outer $\sp$ [SI].
We also find that the radial orbits and low $\sp$ 
emerge from major and minor mergers alike, independent of the 
progenitor mass ratio [SI], and that the presence of a 
$\sim 22\%$ 
bulge does not make a significant difference.  The tidal origin of 
the stellar halo explains this robustness to many merger characteristics.
Furthermore, one simulation continued till 3.5 Gyr 
after the merger with no sign of evolution in  
$\beta(r)$ and $\sp(r)$ beyond $\sim \Re$ [SI].

%15
The $\pm 20\%$ scatter in $\sp$ is partly due to the angular momentum of the
merger orbit and 
%D its inclination relative to the progenitor spins,
the relative spin inclinations, %D
but also due to the line-of-sight 
relative to the principal axes of the triaxial
velocity-dispersion tensor 
%D system %D
(or rotation axis, A. Burkert et al., 
in preparation).  When viewed ``face-on", some of the remnants 
show $\sp$ lower than observed, while other extreme ``edge-on" cases show 
$\sp$ almost as high as 
that of 
the DM [SI].

%16
The simulated line-of-sight velocity distribution 
(LOSVD) 
is consistent with the data also beyond the second moment.  
Both the simulated 
and observed deviations from a Gaussian distribution are typically small, 
%D The deviations from Gaussian are small,
with the fourth moment $h_4 = 0.03 \pm 0.05$ for the central stars [SI].  
At $r\gsim \Re$, radial, prograde mergers produce negligible $h_4$ values 
as in the PNs of NGC 3379\cite{romanowsky03} (or small positive values 
as in NGC 5128\cite{peng04}), while more circular and retrograde mergers,
or gas-rich mergers, 
can produce negative $h_4$ as in NGC 4697\cite{mendez01} [SI].

%%%%%%%%%%%%%%%%%%%%%%%%%%%%%%%%%%%%%%%%%%
\bigskip
\noindent{\bf Conclusion}

%17
\noindent
We conclude that the PN data are consistent with the simple picture
where normal ellipticals also live in massive DM haloes.  
The low $\sp$ is primarily due to the radial orbits of the halo stars, 
being tidally ejected from the inner regions during mergers independently
of dissipation and mass ratio.  This generic origin of the radial orbits
indicates that the results based on our sample of simulations are 
representative of a broader range of merger types expected in the 
$\Lambda$CDM cosmology, and argues that the low observed PN velocities 
are a natural outcome of this standard model. 
This should be confirmed by 
cosmological simulations (e.g.~low-resolution results\cite{saiz04}, and
work in preparation by J. Navarro et al.). 

%18
The range of merger properties leads to a 
variety of $\beta$ and $\sp$ profiles, and the triaxiality adds 
directional variations, allowing extreme $\sp$ values smaller 
and larger than the PN data.
The possible association of the PNs with the younger stars, whose density 
profile is slightly steeper, may help reducing their $\sp$ a bit 
further.  Other tracers, involving old stars, are expected to show a 
somewhat higher $\sp$.  This is especially true for globular 
clusters,\cite{zepf04,napolitano05} 
given their flatter density profile\cite{cote01} and presumably lower 
anisotropy (due to tidal disruption in radial orbits).  
A somewhat higher $\sp$ 
may also be expected in 
gas-poor 
elliptical-elliptical mergers, common especially in groups 
(as observed\cite{napolitano05}), where the collision orbits 
may be circularized by two-body relaxation and dynamical friction. 

%19
Recall\cite{romanowsky03} that a nearly isotropic ``naked" stellar system 
also provides a fit to the $\sp(\rp)$ data inside a few $\Re$, appealing 
to a low $V_0$ rather than a high $\beta$ in \equ{sp}.  
Our simulations provide another solution, which does include a standard 
DM halo.
The DM model predicts that $\sp$ flattens toward $\sim 10 \Re$
(as in NGC 5128\cite{peng04}), where the ``naked" model predicts a 
continuing decline beyond $3\Re$ and very low $\sp$ for other tracers as well.
While the ``naked" model violates much of what we know about galaxy 
formation and cosmology, the DM model, with the radial stellar orbits,
seems to be a straightforward outcome of the self-consistent picture of 
structure formation in the Universe.

%%%%%%%%%%%%%%%%%%%%%%%%%%%%%%%%%%%%%%%%%%%%%%%
%\bigskip
\vfill\eject
\noindent{\bf Methods}

{\small

%----------------------
\medskip\noindent{\bf Equation 2}

\noindent
The coefficient in \equ{sp} is
%D a weak function of $\alpha$ and $\gamma$: %D
%\be
%A(\alpha,\gamma)= \frac{1}{(\alpha+\gamma-1)}
%   \frac{\int_0^{\pi/2} \cos^{\alpha+\gamma}\theta\, d\theta}
%        {\int_0^{\pi/2} \cos^{\alpha-2     }\theta\, d\theta} \ .
%\ee
%It can be expressed in terms of Gamma functions using 
%\be
%\int_0^{\pi/2} \cos^x\theta\, d\theta = \frac{\sqrt\pi}{x}
%{\Gamma\left(\frac{1+x}{2}\right)}/
%     {\Gamma\left(\frac{x}{2}\right)} \ .
%\ee
\be
A(\alpha,\gamma)= \frac{1}{(\alpha+\gamma)}
\frac{\Gamma[(\alpha+\gamma-1)/2]}
     {\Gamma[(\alpha+\gamma  )/2]}
\frac{\Gamma[ \alpha          /2]}
     {\Gamma[(\alpha-1       )/2]}  \ .
\ee
This is a weak function of $\alpha$ and $\gamma$. %D
For instance: $A(3.5,1.0)\simeq 0.18$, 
     $A(3.0,1.0)\simeq 0.20$, $A(3.5,0.4)\simeq 0.23$, $A(3.0,0.4)\simeq 0.26$.

%----------------------
\medskip\noindent{\bf Merger simulations}

\noindent
The 
merger 
simulations\cite{cox04,cox05} represent some of the major 
collisions that likely occurred during the hierarchical structure 
formation according to the $\Lambda$CDM cosmology.  The evolution is 
followed using the entropy-conserving, gravitating, 
Smoothed Particle Hydrodynamics (SPH) code GADGET\cite{springel01_gadget}.  
Gas cooling, star formation and supernova feedback are treated using 
simplified 
recipes that were calibrated to match observed star-formation rates.
% springel02_entropy   springel00_fb 
 
The progenitor disc galaxies mimic typical big spirals galaxies: 
one type (G) representing today's Sb 
galaxies and another type containing more gas as 
in Sbc-Sc galaxies and at high redshift.  The sample consists 
of four G mergers, with DM masses $1.2\times 10^{12}\msun$ (except 
one $5\times10^{11}\msun$), and five Sbc mergers plus one Sc merger
with $8\times 10^{11}\msun$ haloes.  
\ad{REVISED3:}
The baryonic fraction is $\sim 5\%$ of the DM halo mass in the G cases,
and $\sim 13\%$ in the Sbc-Sc cases.  
The fraction of baryons in gas is $\sim 20\%$ in 
G, $52\%$ in Sbc, and 70\% in Sc. 
The particle mass is $<10^6\msun$ for gas and stars and $\lsim 10^7\msun$ 
for DM.  The smoothing lengths are $h=100$ and 400 pc with 
the force becoming Newtonian at $\geq 2.3h$.

Two identical galaxies are set on parabolic orbits and merge because 
of dynamical friction due to their massive haloes.  Our sample consists 
of several different orbits and orientations, including 
prograde and retrograde configurations 
in which the galaxy spins are aligned or antialigned with the 
collision 
orbital angular momentum.  
The merger results in two succesive 
starbursts, one after the first close approach, and the other 
after the second, final coalescence [SI].  The starbursts 
typically 
occur 1-2 Gyr after the beginning of the simulation, and the 
remnant is ``observed" $\sim 1$ Gyr later.  The amount of 
stars formed during the merger is roughly proportional to the 
initial gas fraction, and is not too sensitive to the 
orbit or orientation.  The instantaneous rate 
varies in the range 
%D is %D
10-100 $\msun$/yr.  The young stars formed during the merger constitute 
$\sim 30\%$ of the total stars; typically 20\% in the G 
cases and 40\% in Sbc. 
The remnant galaxies resemble normal elliptical galaxies, as 
demonstrated above [SI].

%D From SI
The smaple used from the suite of 1:1 merger
simulations\cite{cox04,cox05} consists of the following runs:

The G cases are:
G3G3b-u1 (the fiducial G case,
a gas-poor galaxy obeying the scaling relations from SDSS),
G3blv5G3blv5b-u1 (bulgeless),
G3G3r-u1 (a retrograde orbit),
G2G2r-u1 (a slightly smaller gas-poor galaxy obeying the SDSS scalings,
on a retrograde orbit).

The Sbc-Sc cases are:
Sbc201a-u4 (the fiducial Sbc case, gas rich, small bulge,
same as run n2low\cite{cox05}),
Sbc201-uu43 (with stronger feedback, n2high\cite{cox05}),
Sbc203-u4 (retrograde orbit),
Sbc204-u4 (very radial orbit),
Sbc205-u4 (more tangential orbit),
Sc201-u4 (mimicking Sc galaxies with no bulge and a higher gas fraction).

%--------------------------------
\medskip\noindent{\bf Scaling the data}
 
\noindent
The observed galaxies are presented together using the $\Re$
determined from 
each surface-brightness profile\cite{devac79,peletier90,romanowsky03}.
We note that $\Re$ for NGC 3379\cite{devac79,peletier90} is 
$\simeq 50\%$ larger than that quoted\cite{romanowsky03}. %burstein87
In \fig{2d}, an open circle marks the last point had we used 
the smaller\cite{romanowsky03} $\Re$; 
this uncertainty does not make 
a qualitative difference.  The amplitudes of $\Sigma$ and $\sp$ are scaled by 
least-squares fits of the stellar data at $r>0.2\Re$ to 
the stacked simulated profile as a reference. Replacing this reference 
by a different function of a similar general shape yields similar results. 
Using only the stars at larger radii (up to $r>\Re$), or using the PNs 
alone, yield $\sp$ adjusting factors 
that differ only by a few percent.

The $\Re$ of NGC 821, 3379, 4494, 4697 match the mean simulation
value after multiplication by 
      0.667, 1.57, 1.00, 1.13,
indicating that the simulated and observed galaxies are of comparable sizes.
The $\sp$ were adjusted by factors 
1.00, 1.19, 1.21, 1.11 
for best fit.  Being comparable to the radius scaling factors indicates
that the observed and simulated galaxies have a similar
velocity structure. The mean and $1\sigma$ scatter in the
simulated remnants are $\Re=4.05\pm 1.04 \kpc$ and $\sp(\Re)=154\pm 33 \kms$.

%----------------------
\medskip\noindent{\bf Age of Planetary Nebulae}

\noindent
The [O{\small III}] luminosity of a PN with 
mass $<2.5\msun$ is strongly increasing with the parent stellar 
mass\cite{marigo04} [Figs.~10, 14], hence sharply decreasing with age. 
A limiting magnitude $M_{5007}$ then corresponds to a maximum stellar age $t$.
For the complete sample of 328 PNs in NGC 4697\cite{mendez01} it 
is $M_{5007}\simeq -2.6$, namely $t_{M01}<3$ Gyr\cite{marigo04} 
[Figs.~18, 19, 26].  With only $\simeq 100$ 
PNs per galaxy\cite{romanowsky03}, the magnitude limit is brighter (by -0.8 
magnitudes based on telescope gathering areas), so the stars are 
even younger.  Based on theoretical PN luminosity 
functions\cite{marigo04} [Figs.~18, 26], 
if the population is typically older than 1 Gyr then $t_{R03}<2$ Gyr. 
We therefore adopt $t<3$ Gyr as a limit for most 
of the observed PNs in the four galaxies. This indicates 
an association with the ``young" simulated stars, and that the mergers 
of gaseous discs are relevant to those ellipticals showing PNs.  
A caveat is the apparent relative invariance of the PN 
luminosity function between galaxies, seemingly independent of signs for a 
recent major merger.  When there are no 
signs for a recent major merger, 
the observed PNs, 
if indeed young, 
may be the signature of recent minor mergers, which
are expected to produce similar effects.

%%%%%%%%%%%%%%%%%%%%%%%%%%%%%%%%%%%%%%%%%%%%%%%%%%
\vfill\eject

\addtolength{\baselineskip}{-0.05\baselineskip}
\addtolength{\parskip}{-0.2\baselineskip}

\bibliographystyle{natureedo}
\bibliography{pn_astroph}
%\bibliography{pn}

\addtolength{\baselineskip}{+0.05\baselineskip}
\addtolength{\parskip}{+0.2\baselineskip}

%%%%%%%%%%%%%%%%%%%%%%%%%%%%%%%%%%%%%%%%%%%%%%%
\vskip 0.5cm
\noindent{\bf Acknowledgments}
% Revised
We acknowledge stimulating discussions with M. Beasley, A. Burkert,
K. Gebhardt, J. Navarro, A. Romanowsky and his group\cite{romanowsky03},
and assistance from M. Covington.
This research has been supported by 
ISF 213/02 and by NASA ATP NAG5-8218.
AD acknowledges a Miller Professorship at UC Berkeley, 
support from UCO/Lick Observatory, and a Blaise Pascal 
International Chair in Paris.

\vskip 0.5cm
\noindent{\bf Competing interests statement} The authors declare that 
they have no competing financial interests.

\vskip 0.5cm
\noindent {\bf Correspondence} and requests for materials should be 
addressed to A.D. (dekel@phys.huji.ac.il).
%%%%%%%%%%%%%%%%%%%%%%%%%%%%%%%%%%%%%%%%%%%%%%%%%%

\vfill\eject
\centerline{\bf SUPPLEMENTARY INFORMATION}
\bigskip

%%%%%%%%%%%%%%%%%%%%%%%%%%%%%%%%%%%%%%%%

\no
The following is an extension of the Letter to Nature,
aimed at providing more detailed support to the results reported 
in the Letter.

\smallskip
\no
{\bf Outline:}
\begin{enumerate}
  \item The generic origin of a stellar halo with radial orbits
  \item Robustness to dissipation
  \item Robustness to major versus minor mergers
  \item Robustness to the presence of a bulge
  \item Long-term stability
  \item Parametrizing the velocity-dispersion profiles
  \item The full LOSVD and the $h_4$ moment
  \item Matching other properties of ellipticals
  \item Scatter induced by triaxiality
  \item On the PN age
  \item Rising versus constant $\beta$
  \item Why was the DM solution missed before?    
  \item Robustness and Uniqueness of the results
\end{enumerate}

\Fig{snaps_sbc} shows snapshots from one of our merger simulations, 
brought here
for general impression of some of the issues discussed in the Letter
and in this SI.

\begin{figure}[t] %1
\vskip 8.8cm
{\includegraphics{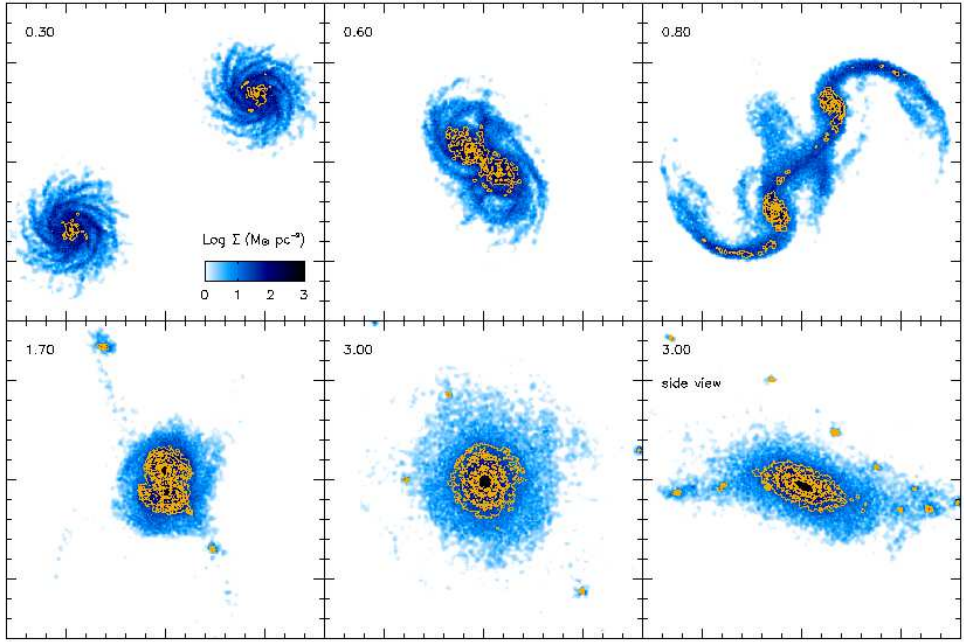}}
\vskip 0.2cm
\caption{Visual impression. 
Stellar density maps in one of the merger simulations (fiducial Sbc).
``All" the stars (blue) and the ``young" stars formed during the merger
(yellow contours).
The time sequence (time marked in Gyr) is viewed ``face-on"
in the plane defined by one of the initial discs and the two centers of mass.
Young stars form especially after each of the two pericentres, at
$\sim 0.60$ and $\sim 1.70$ Gyr after the beginning of the simulation.
The final system is also viewed ``edge-on" (bottom-right),
showing an axial ratio of roughly 1:2.
The ``young" stars have a somewhat steeper density profile
and a slightly higher ellipticity.
}
\label{fig:snaps_sbc}
\end{figure}

%----------------------------------
%1
\section{The generic origin of a stellar halo with radial orbits}

\begin{figure}[t] %2 TJ  
\vskip 8.8cm
%{\special{psfile="contour0.ps"  hscale=37 vscale=37 hoffset=-7 voffset=37}}
%{\special{psfile="contour12.ps" hscale=37 vscale=37 hoffset=120 voffset=37}}
%{\special{psfile="contour20.ps" hscale=37 vscale=37 hoffset=247 voffset=37}}
%{\special{psfile="contour36.ps" hscale=37 vscale=37 hoffset=-7 voffset=-90}}
%{\special{psfile="contour44.ps" hscale=37 vscale=37 hoffset=120 voffset=-90}}
%{\special{psfile="contour60.ps" hscale=37 vscale=37 hoffset=247 voffset=-90}}
{\includegraphics{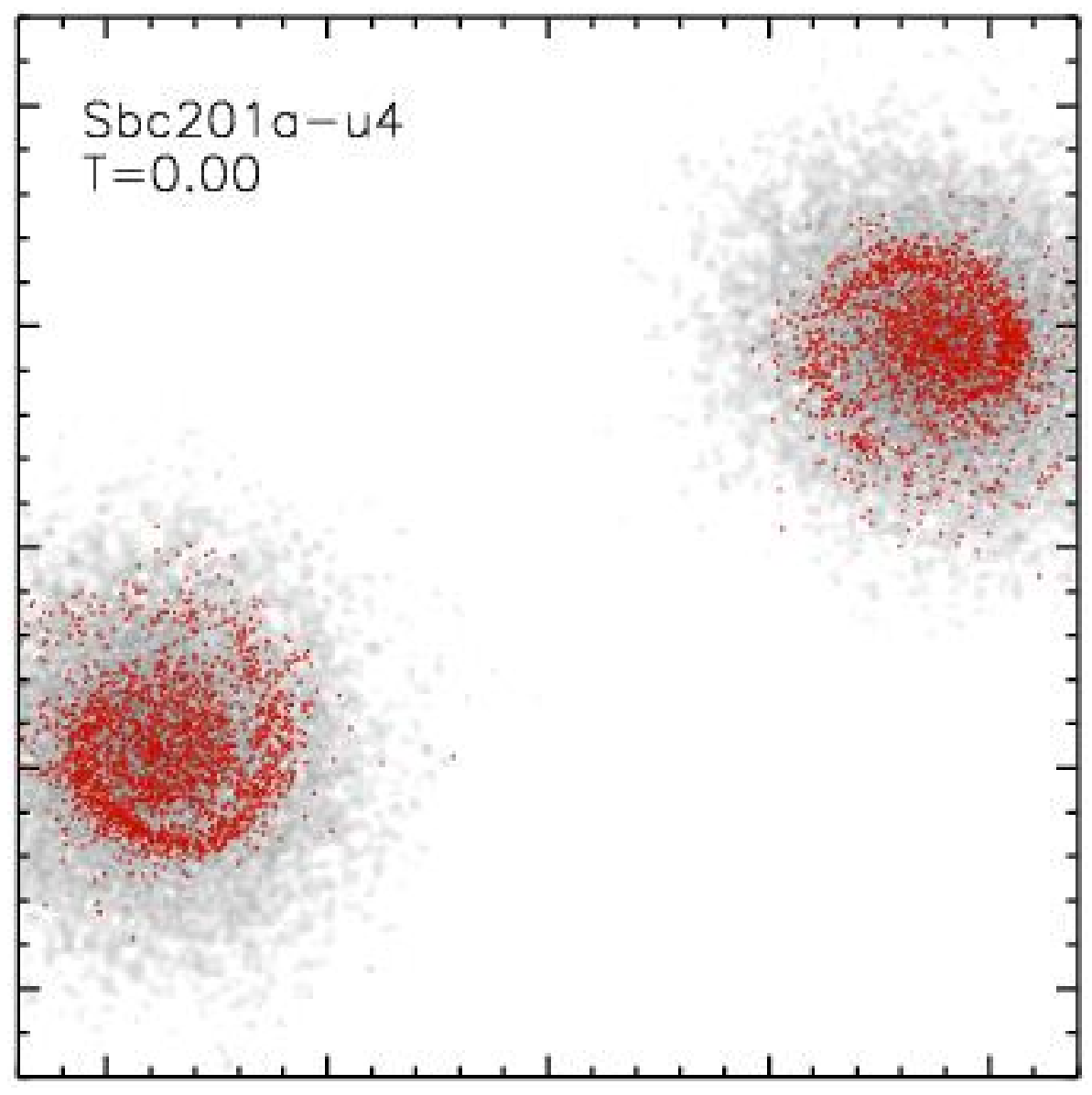}}
{\includegraphics{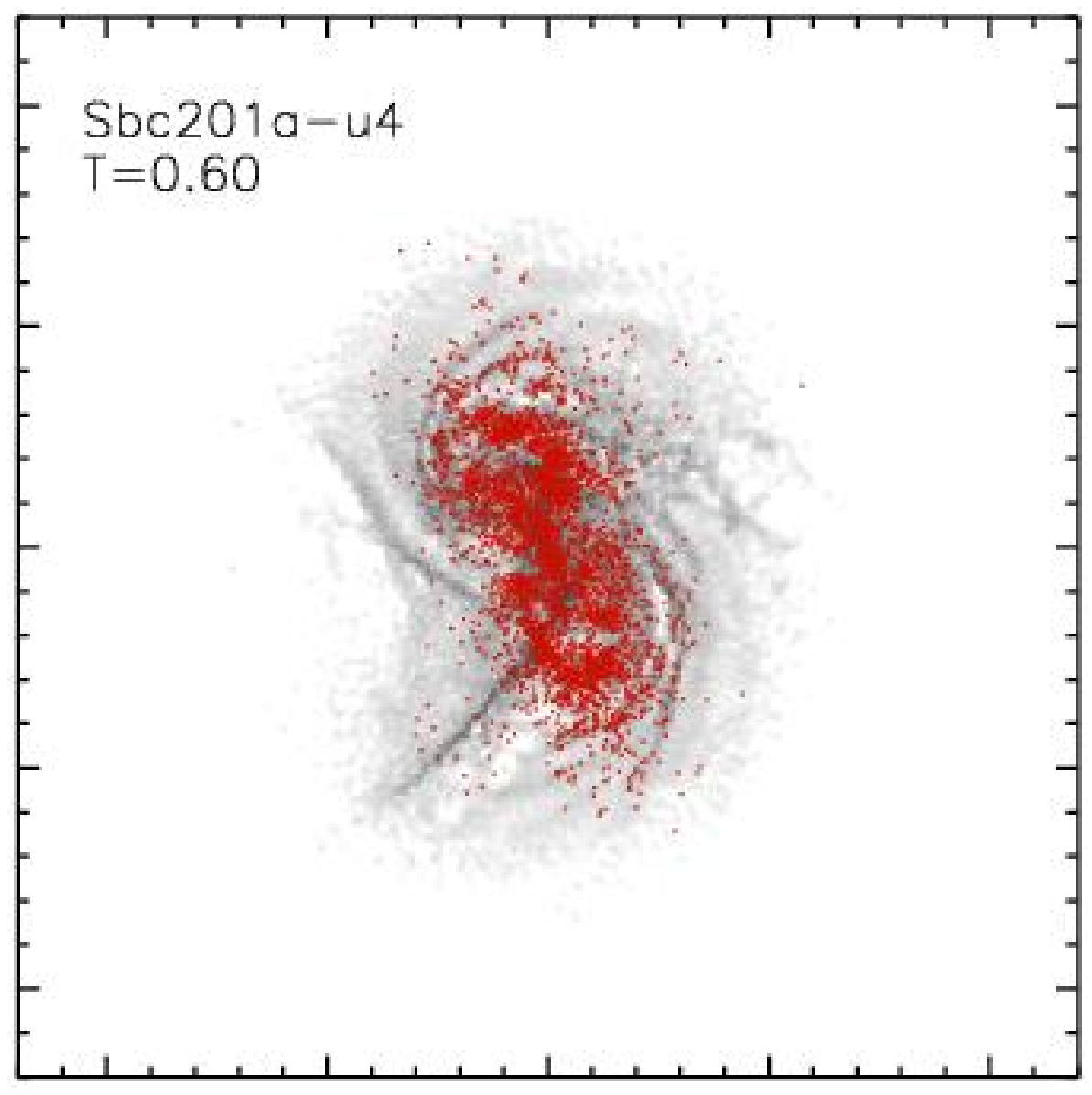}}
{\includegraphics{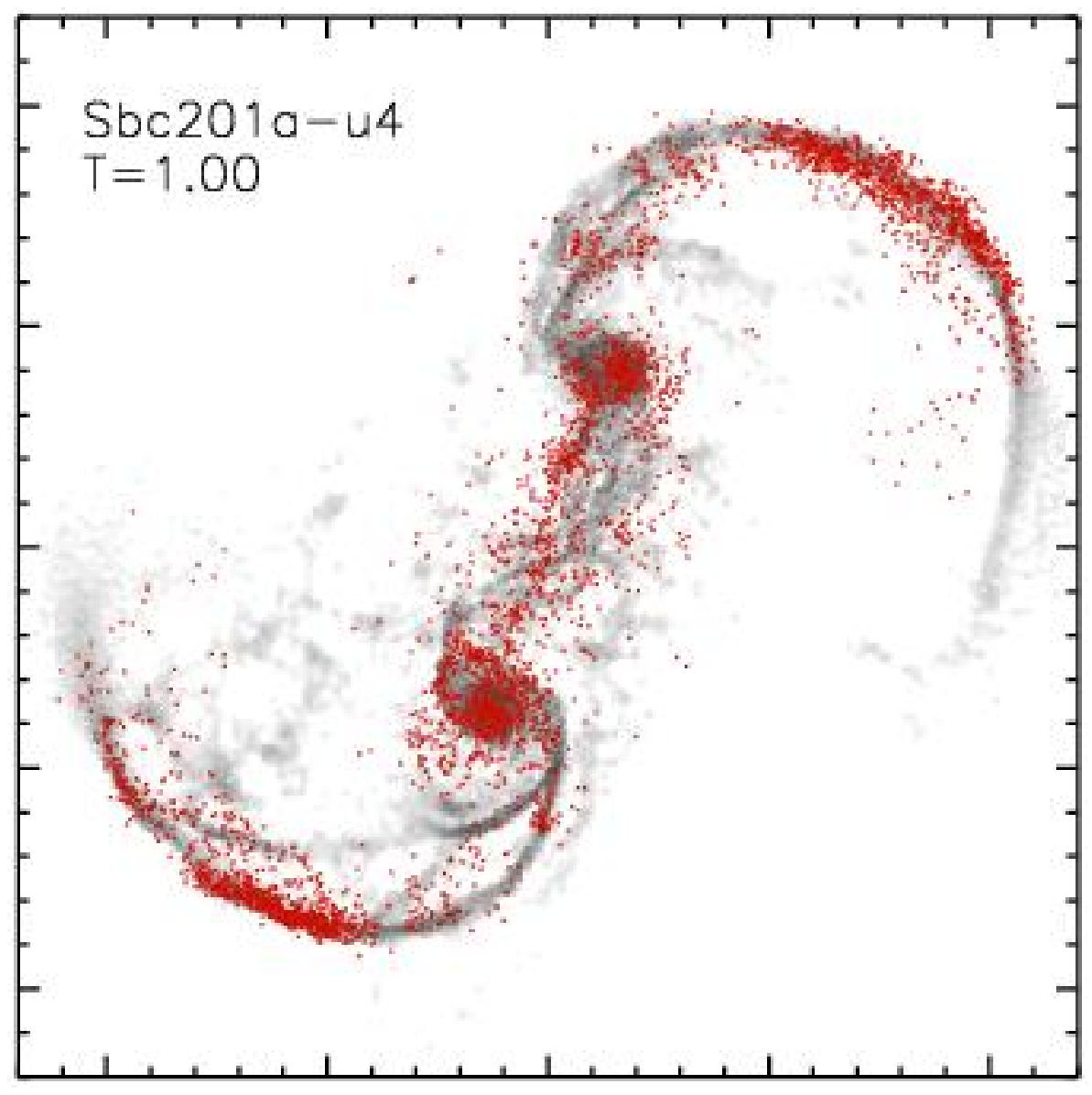}}
{\includegraphics{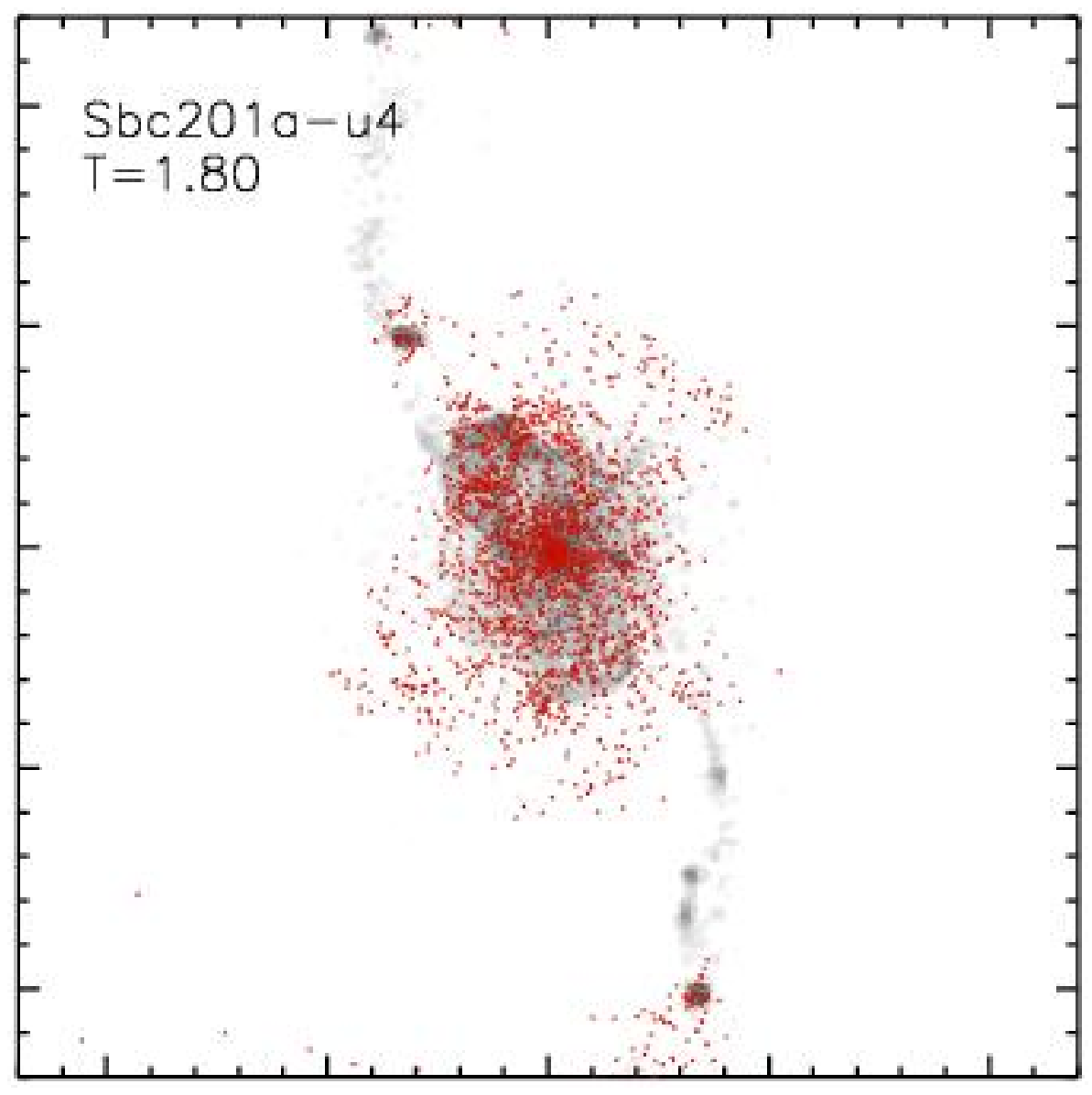}}
{\includegraphics{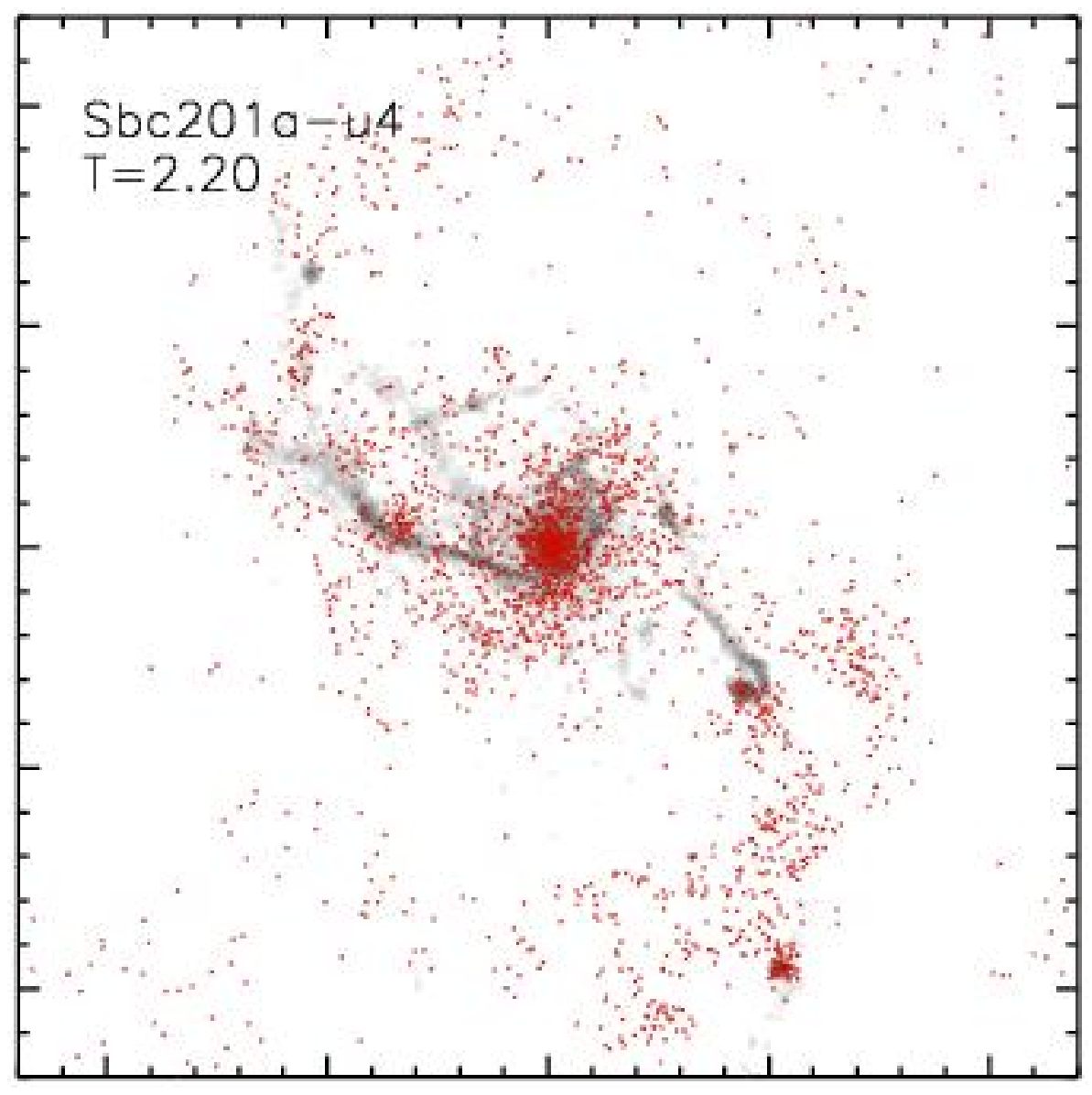}}
{\includegraphics{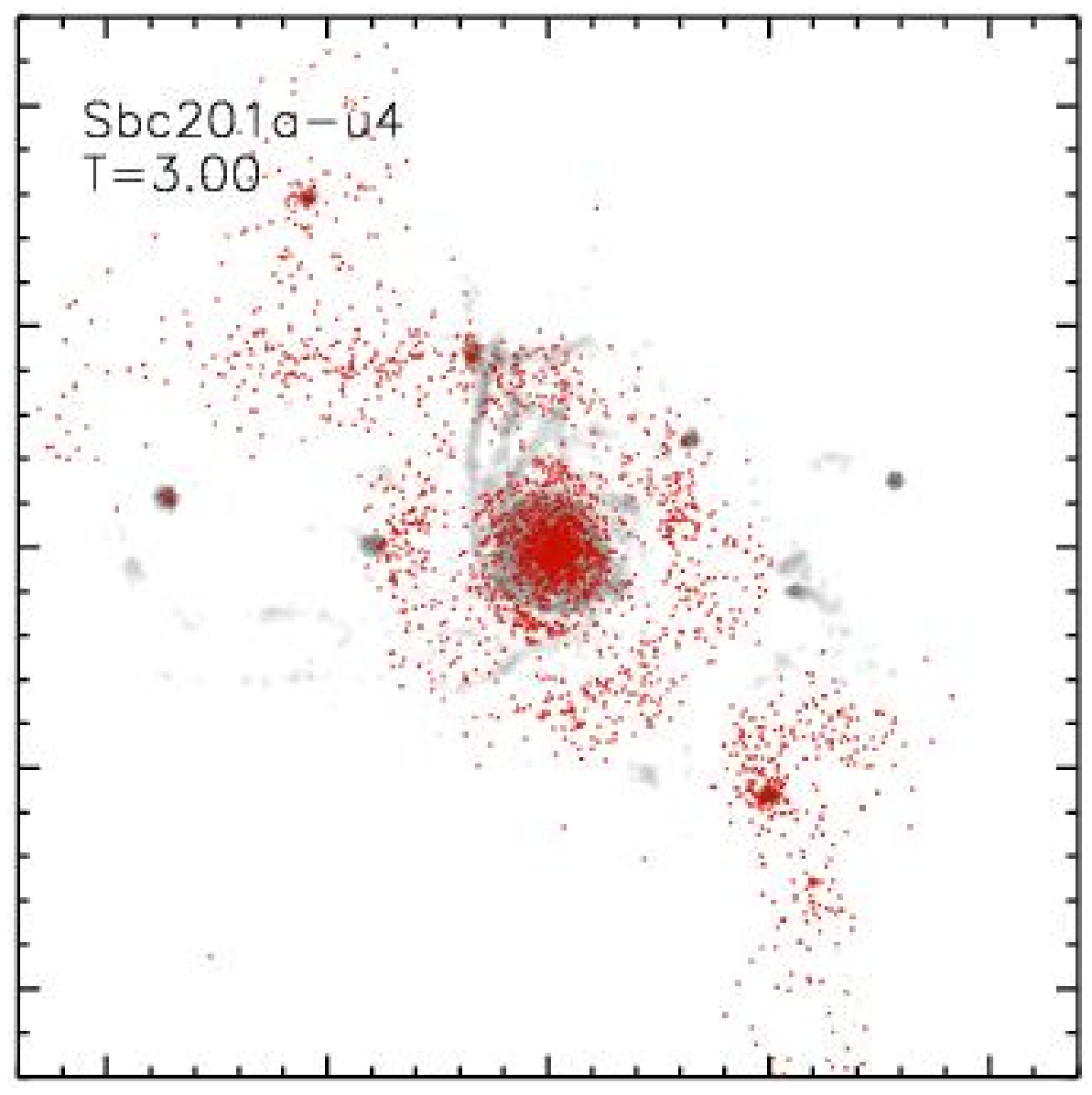}}
\vskip 0.2cm 
\caption{The tidal origin of the stellar halo and radial orbits is demonstrated
via Snapshots from the fiducial Sbc simulation (same as \fig{snaps_sbc}).
Marked red is a sample of (``old") stars identified at the final time ($t=3$)
as loosely bound.  These stars are shown in their positions at several
stages along the merger process. 
The final $\Re$ is about two small tick marks.
At the initial time, these stars are spread throughout the
progenitor disks ($t=0$).
The first pericentre passage is at $t \simeq 0.6$, and
during the rebound ($t =1.0$) 
these stars are divided into two groups: 
those that are flying outward with the extended tidal tails, and those that
are still associated with the progenitor cores as well as the tidal bridge
between them.
At the second-pericentre coalescence ($t \simeq 1.8$), the outer 
tidal tails are not seen while the inner population is expanding from the
center into the halo on highly radial orbits.
At $t=2.2$ we see the stars from the tidal tails falling back in
on highly radial orbits, to eventually join the inner group
in a loosely bound stellar halo extending to a few $\Re$.
}
\label{fig:radial_origin}
\end{figure}

\fig{radial_origin} demonstrates the important role of {\it tidal\,}
processes in the formation of a stellar halo and its radial orbits.
The final halo stars are composed of two populations:
(a) stars that passed through the center near the second-pericentre
coalescence, some of which were part of the tidal bridge formed after the
first-pericentre passage,
and (b) stars that first flew outward in extended
tidal tails after the first-pericentre passage and then
fell back after the merger of the cores.

The tidal origin of the radial orbits explains 
the robustness to the degree of dissipation, the progenitor
mass ratio, the bulge fraction and so on.
It predicts that $\beta$ should predominantly be a function of the strength
of the tidal interactions associated with the collision. 
Indeed, we find in our simulations a strong tendency for high $\beta$ 
in collisions of low impact parameter, as well as in prograde collisions,
where the tidal effects are stronger. 
High angular momentum collisions, and retrograde
collisions, typically lead to low beta values.
Such collisions may be responsible for elliptical galaxies with higher velocity
dispersions than the ones discussed in the Letter.
Future work may try to understand how such collisions are associated
with bigger ellipticals preferentially in group centers (e.g.
due to two-body relaxation and orbit circularization by dynamical friction).

How sensitive are our results to the choice of parabolic orbits in the
simulations? 
We extracted our typical orbits from cosmological simulations\cite{khochfar03}.
We have performed a few tests of different orbits and orientations,
but have not obtained conclusive results about the trends.
What we can say at this point is that the sample of simulations that
we have used spans a range of properties that matches many of the real
galaxy properties, and especially reproduces many cases of low $\sp$.
The dependence on orbit is worth further study.

%-----------------------------
%2
\section{Robustness to dissipation}

We find that the role of dissipation on the orbit anisotropy beyond $\Re$
is negligible. The main result of dissipation is a more centrally concentrated
stellar distribution, which results in $\sim 10\%$ reduction in
velocity dispersion -- a minor effect. 

One demonstration is provided by the comparison of the ``young" stars, 
involving dissipation, to the dissipationless ``old" stars. 
The $\beta$ profiles are similar (Fig.~1 of the Letter) 
and the $\sp$ profiles differ by only $\sim 9\%$ at $3\Re$ (Fig.~2 of the
Letter).

A further demonstration is provided in \fig{dry}, which compares 
the results for three merger simulations with
the initial gas/baryon fraction ranging from 0 to 70\%.
In particular, we added to our original sample a case identical to
our fiducial Sbc merger (52\% gas) but with all the gas particles treated
as non-dissipative particles.
This ``dry" case ends up with a slightly {\it higher\,} $\beta$ (!), which is
balanced by a slightly lower $\alpha+\gamma$, resulting in a slightly 
higher $\sp$.

\begin{figure} %3  Felix
\vskip 7.0cm
%{\special{psfile="dry_beta.eps" hscale=38 vscale=38 hoffset=10 voffset=-13}}
%{\special{psfile="dry_sp.eps" hscale=38 vscale=38 hoffset=230 voffset=-13}}
{\includegraphics{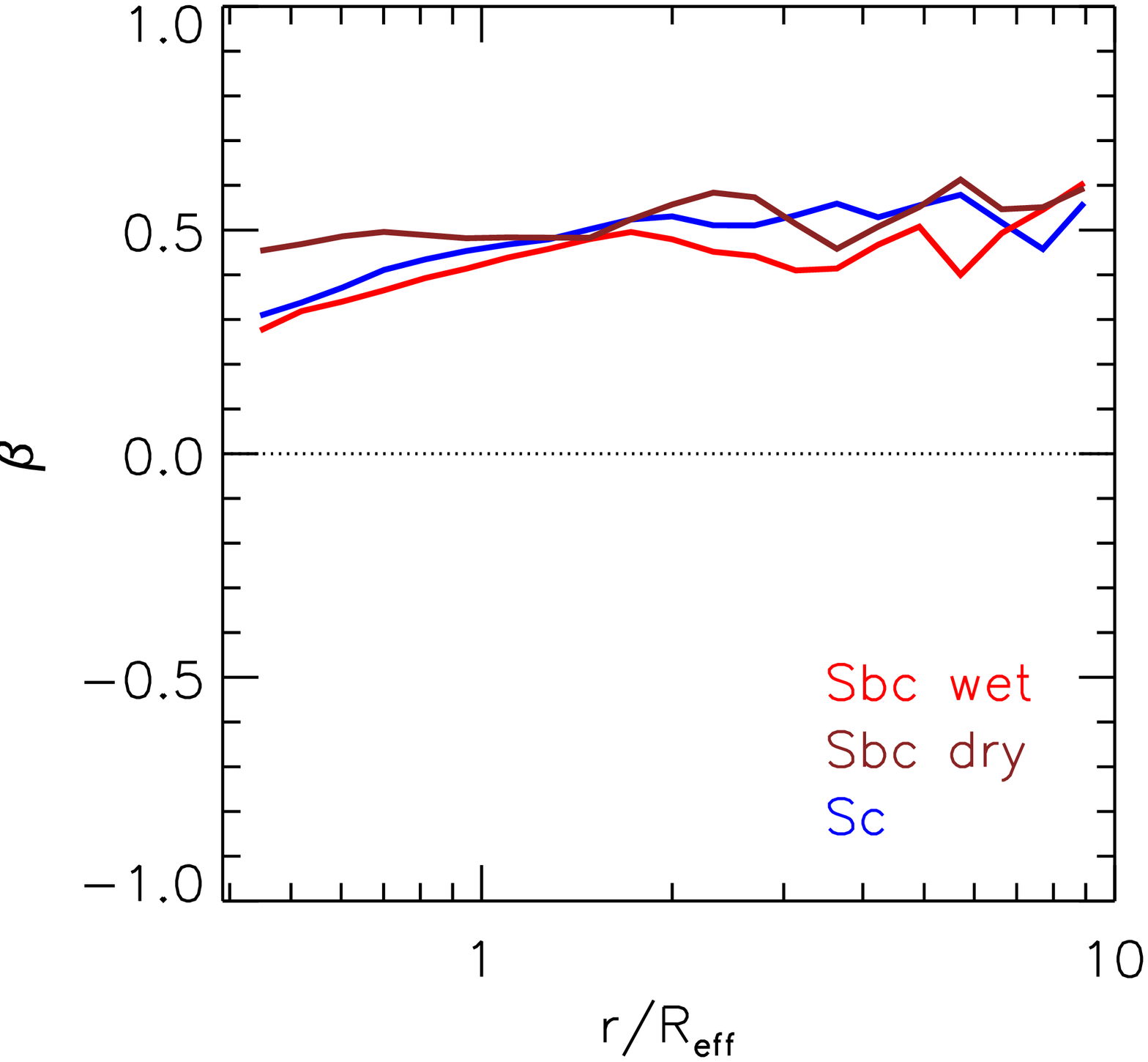}}
{\includegraphics{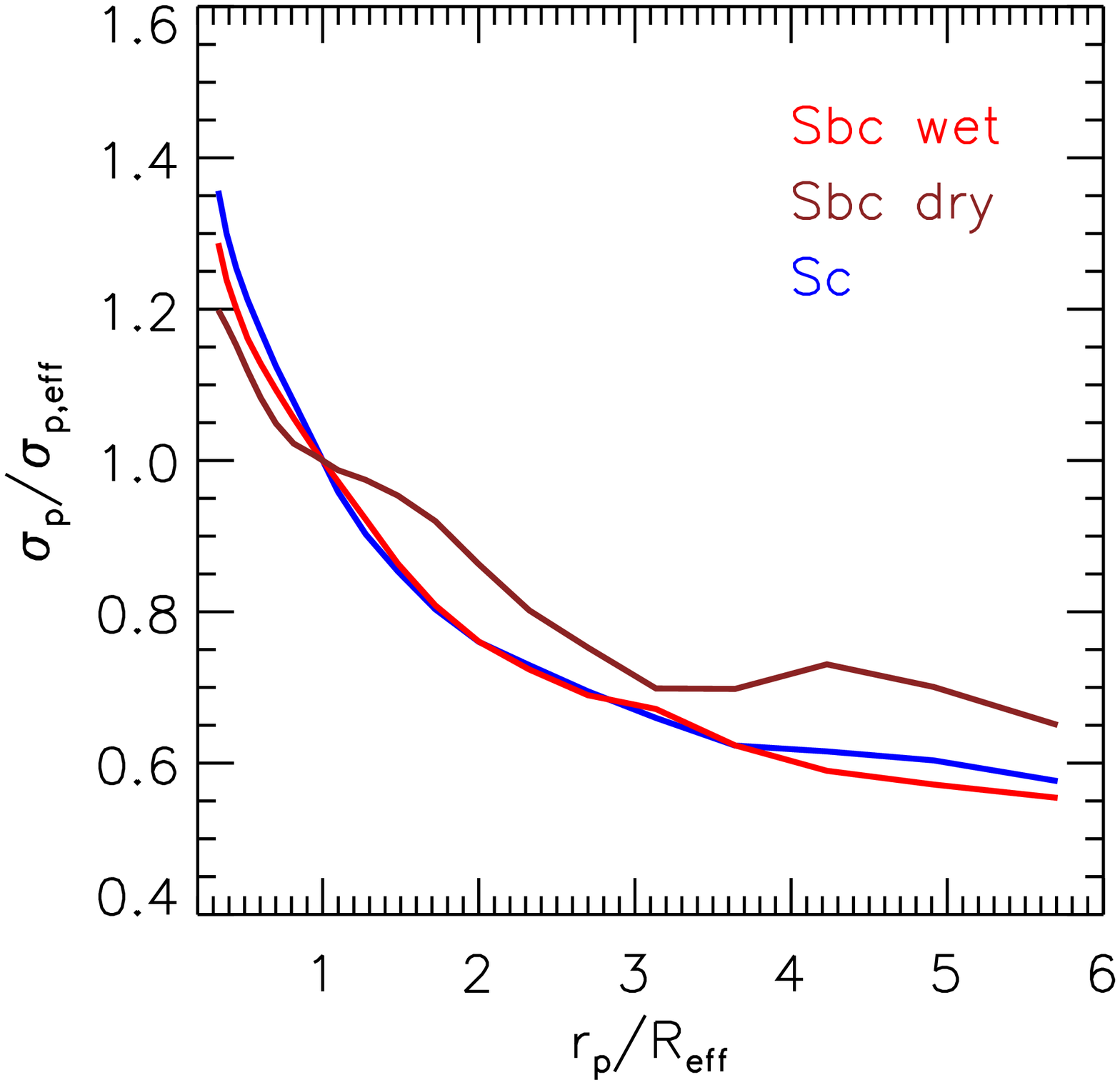}}
\vskip 0.2cm
\caption{Robustness to dissipation.
Anisotropy and velocity dispersion of the ``old" stars 
for mergers with different gas fractions.
Three projections at two time steps are stacked at $\Re$.
The red and brown correspond to the same Sbc merger with initially 52\%
and zero gas/baryons respectively.
The blue is an Sc merger with 70\% initial gas.
The effect of dissipation is small, at the level of $\sim 10\%$ in $\sp$.
}
\label{fig:dry}
\end{figure}

We conclude that the radial orbits are generic to mergers independent of
whether they are ``wet" or ``dry". The effect of dissipation on $\sp$, via its 
effect on the density profile, is limited to $\sim 10\%$,  
and is therefore of only secondary concern. 
This indicates that the key results from our merger sample are
representative of a variety of mergers, ``wet" and ``dry", expected in a
LCDM universe. 

One could argue in passing that the higher $\alpha+\gamma$ resulting from 
the ``wet" mergers, being closer to the observed $\alpha+\gamma$ in the 4 
ellipticals discussed (based on Fig.~2 of the Letter), indicates that some 
degree of dissipation is actually required in order to produce the correct
density and velocity profiles of elliptical galaxies. This is, however, 
only a marginal indication which is worth further study.    

The effects of varying the recipes for star formation and feedback
on the relevant features addressed in this Letter are yet to be explored.

%------------------------------------
%3
\section{Robustness to major versus minor mergers}

\begin{figure}[t] %4  Felix
\vskip 7.0cm
%{\special{psfile="small_beta.eps" hscale=38 vscale=38 hoffset=10 voffset=-13}}
%{\special{psfile="small_sp.eps" hscale=38 vscale=38 hoffset=230 voffset=-13}}
{\includegraphics{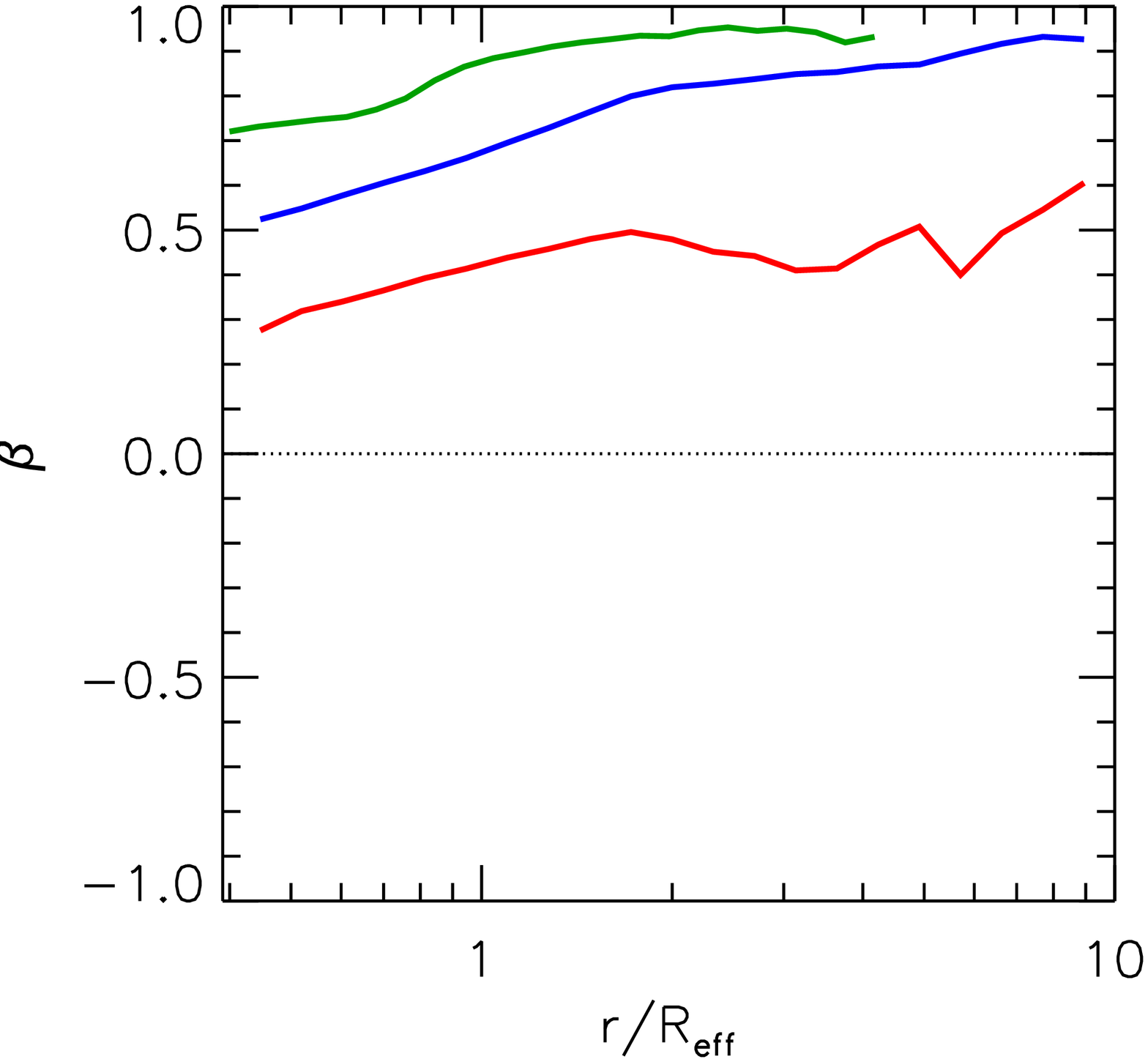}}
{\includegraphics{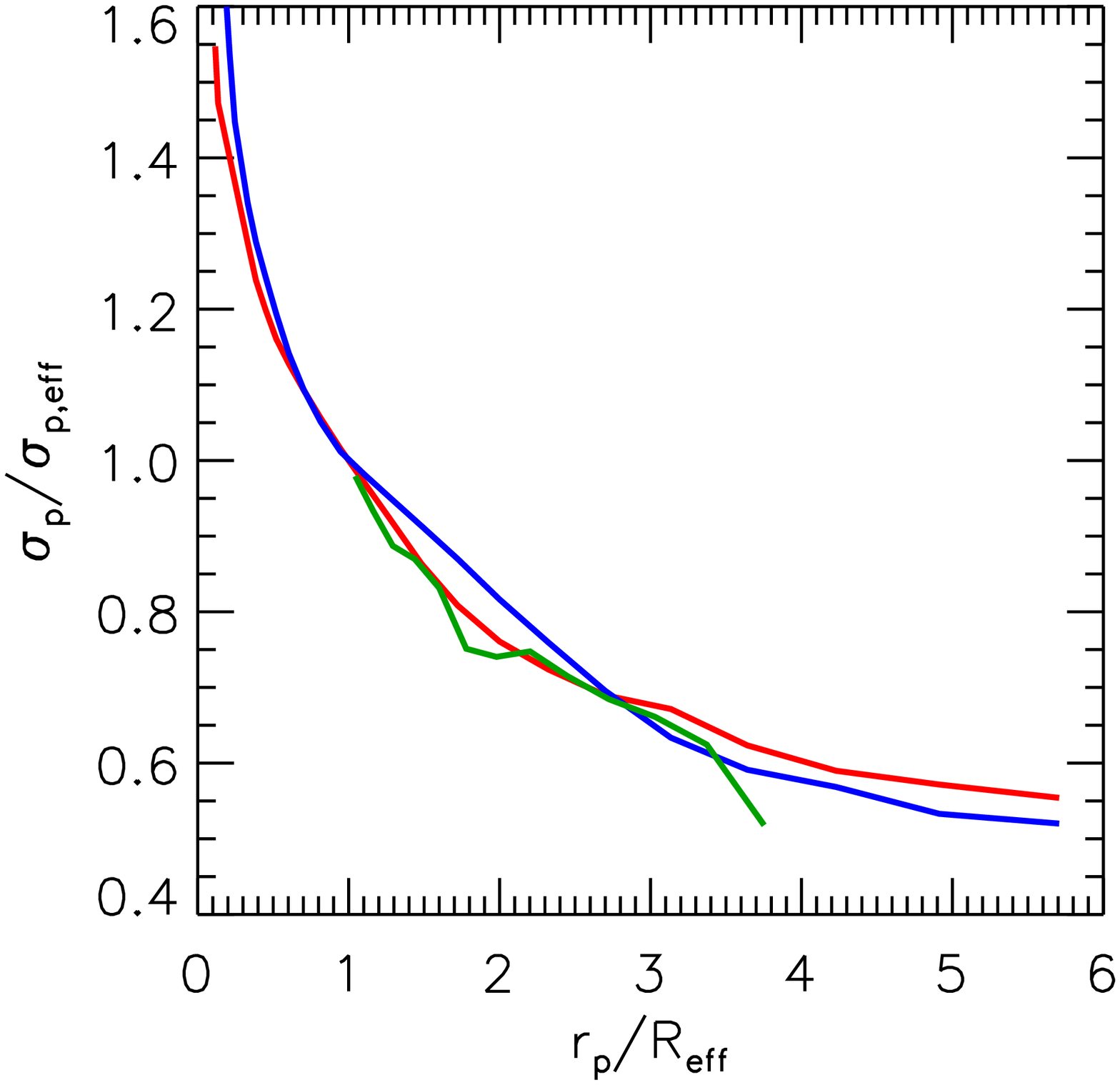}}
\vskip 0.2cm
\caption{Robustness to major versus minor mergers.
Anisotropy and velocity dispersion for major and minor
mergers with different mass ratios.
For each merger, three projections at two time steps are stacked at $\Re$.
Red: our fiducial 1:1 Sbc merger.
Blue: four 1:3 G mergers stacked together. 
Green: four 1:10 minor mergers stacked together.
In the unequal mass mergers, only the ``old" stars of the smaller progenitors 
are used, to better mimic the halo population after a sequence of such mergers
and avoid the surviving disk component of the bigger progenitor.
The orbits tend to be more radial for the unequal mass mergers, while
the final dispersion profiles are hardly distinguishable from the 1:1 case.
}
\label{fig:minor}
\end{figure}

Our key results are not restricted to 1:1 mergers either.
This is demonstrated in \fig{minor}, which compares our fiducial 1:1 case
with several 1:3 and 1:10 mergers.
After a single minor merger, the remnant does not necessarily resemble
an elliptical galaxy (e.g. the disk of the big progenitor is still intact),
but the halo stars originating in the small progenitor are on very 
radial orbits, yielding final dispersion profiles which 
are hardly distinguishable from the 1:1 case.
The radial orbits and the low velocity dispersion
are thus generic to the stellar haloes of merger remnants, major or minor.
This confirms what we heard privately from J. Navarro regarding their
preliminary findings based on a cosmological sequence of mergers.
It indicates that the key results from our merger sample are
representative of a variety of mergers, major and minor, expected in a
LCDM universe. 

Our methodology for checking the effects of
minor mergers is limited by the fact that
we have performed so far single mergers only. 
Our logic is that after multiple minor mergers,
the final profile is dominated by the many small progenitors 
which are all contributing to increased radial anisotropy. 
The detailed study of minor mergers is beyond the scope of this Letter.

%-----------------------------
%4
\section{Robustness to the presence of a bulge}

\begin{figure} %5  Felix
\vskip 7.0cm
%{\special{psfile="bulge_beta.eps" hscale=38 vscale=38 hoffset=10 voffset=-13}}
%{\special{psfile="bulge_sp.eps" hscale=38 vscale=38 hoffset=230 voffset=-13}}
{\includegraphics{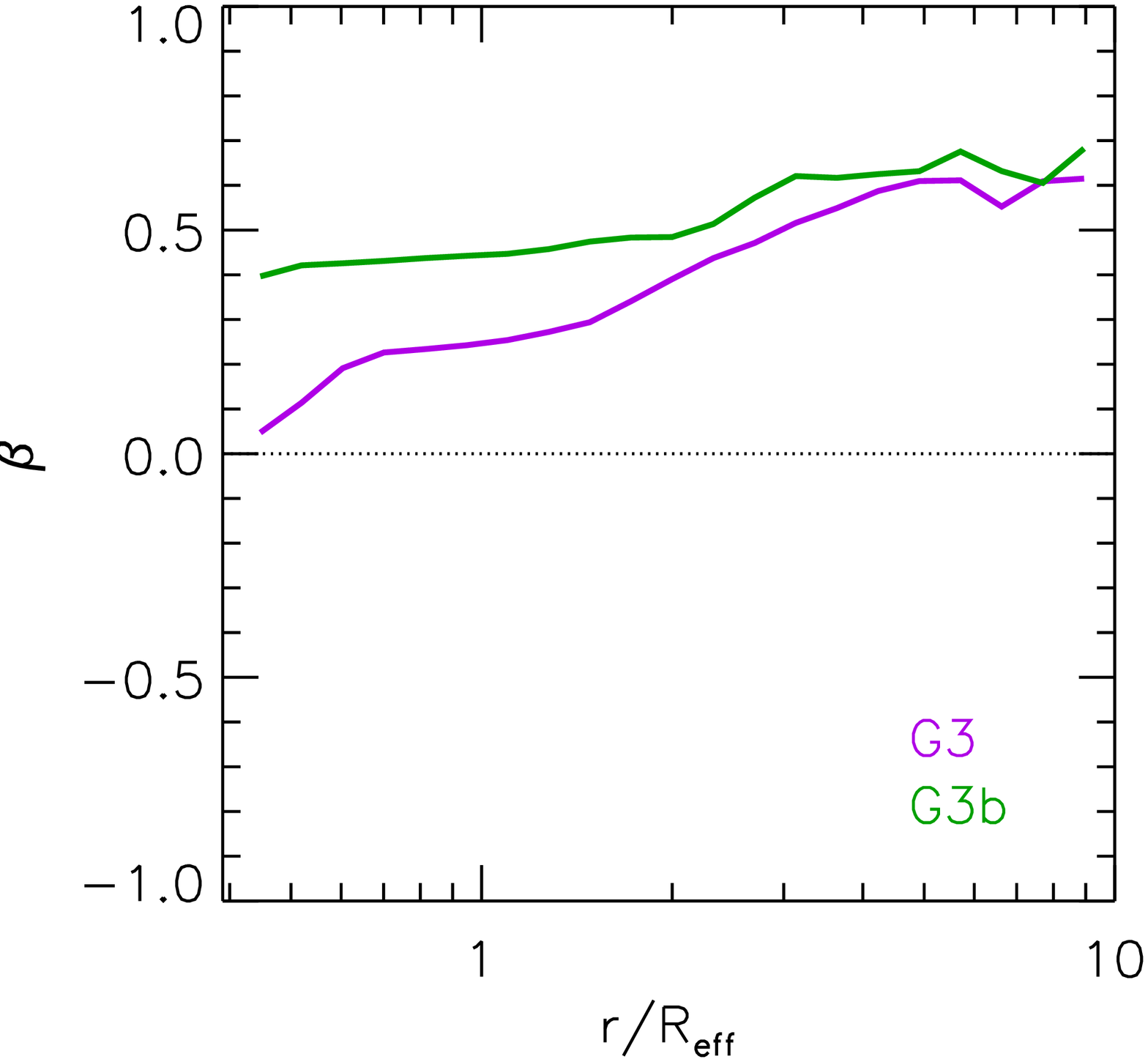}}
{\includegraphics{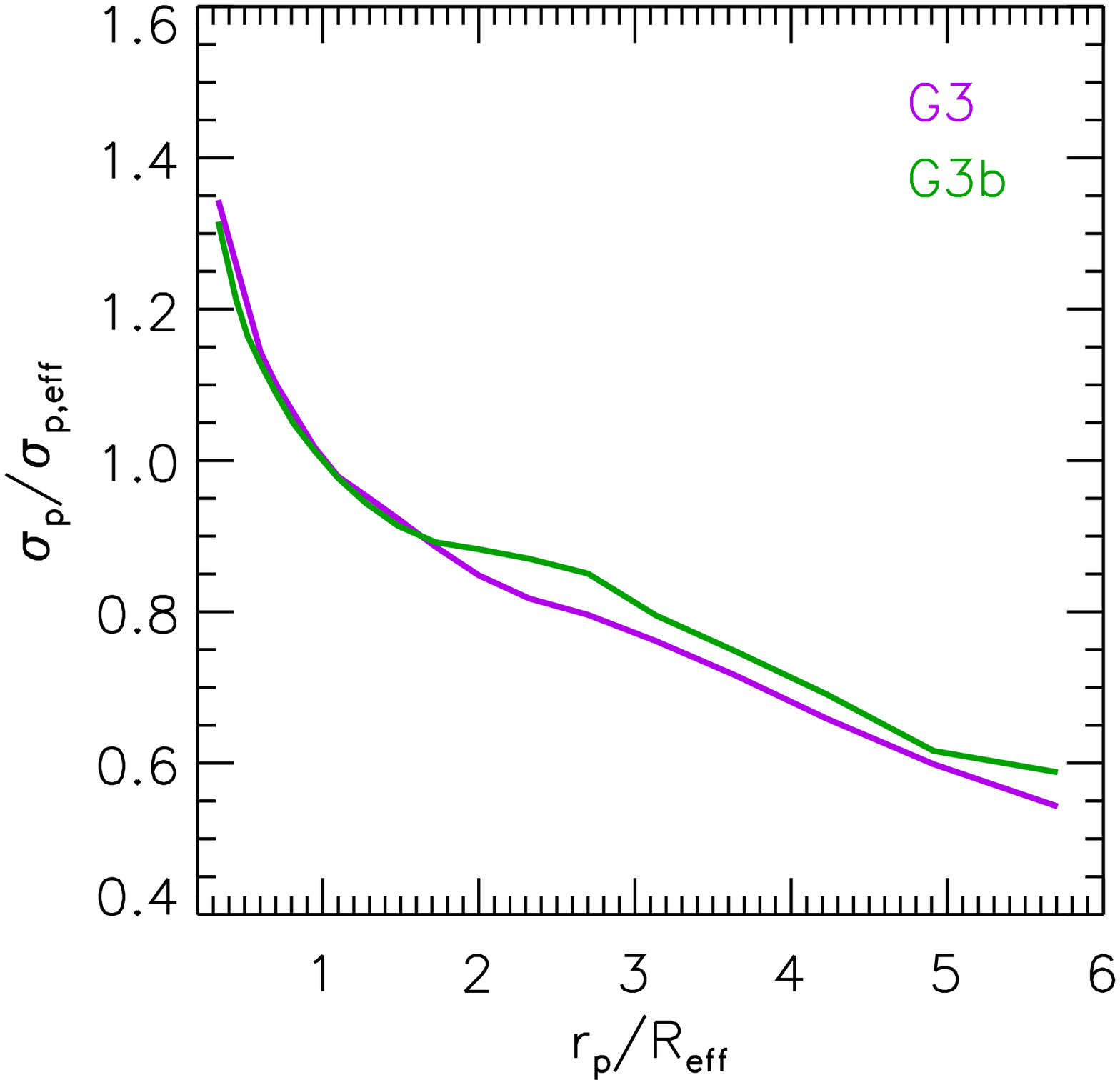}}
\vskip 0.2cm
\caption{Robustness to the presence of a bulge.
Anisotropy and velocity dispersion of the ``old" stars
for mergers with and without a $22\%$ bulge component.
Three projections at two time steps are stacked at $\Re$.
The purple is a G merger with 20\% initial gas to stars+gas 
and 22\% bulge to stellar disk ratio.
The green is the same but excluding the bulge.
}
\label{fig:bulge}
\end{figure}

The effect of a relatively small bulge is seen in \fig{bulge}.
It reduces $\beta$, but increases the central stellar concentration,
together leaving the velocity dispersion hardly changed.

We cannot tell from our current simulations what happens when
two ellipticals merge.  We may speculate that
if $\beta$ turns out to decrease in such mergers (as indicated by the
effect of a bulge), while the density profile does not steepen (unlike the
case of a bulge), the result may be a slight increase in $\sp$.
Given that galaxies in groups are more likely to be ellipticals
than in the field\cite{helsdon03b},
the dominant group elliptical is more likely to be formed by the mergers
of elliptical galaxies (or bulge-rich spirals). If $\beta$ is indeed
lower in the corresponding remannts, and if the density profiles are
the same as in spiral mergers, one would predict that dominant
group members, which are also the most massive ellipticals,
should have higher dispersions, as is indirectly observed\cite{napolitano05}.
This will be interesting to address in future work, but it is not required
for the main point of our current Letter concerning the puzzling
low-dispersions in intermediate ($L_*$) ellipticals.

% (e.g. Helsdon & Ponman 2003, MNRAS 339, L29), and

%------------------------
%5
\section{Long-term stability}

We have continued one of the G mergers for a longer time, for a total
of 6Gyr, where the merger occurred at $\sim 2.5$ Gyr.
Shown in \fig{time} is the time evolution of the profiles of
$\beta$ and $\sp$. During the first 0.75 Gyr after the merger
(3 dark curves), the system is not yet fully relaxed, though the high $\beta$
and low $\sp$ are already established.
During the following 3 Gyr (blue and light-blue curves) the profiles are
very stable, at least until 3.5 Gyr after the merger.
The slight evolution is limited to
central regions where remnant gas is slowly forming more stars.

\begin{figure}[t] %6
\vskip 7.6cm
%{\special{psfile="beta_time.eps" hscale=68 vscale=68 hoffset=0  voffset=-10}}
%{\special{psfile="sigma_time.eps" hscale=68 vscale=68 hoffset=235 voffset=-10}}
{\includegraphics{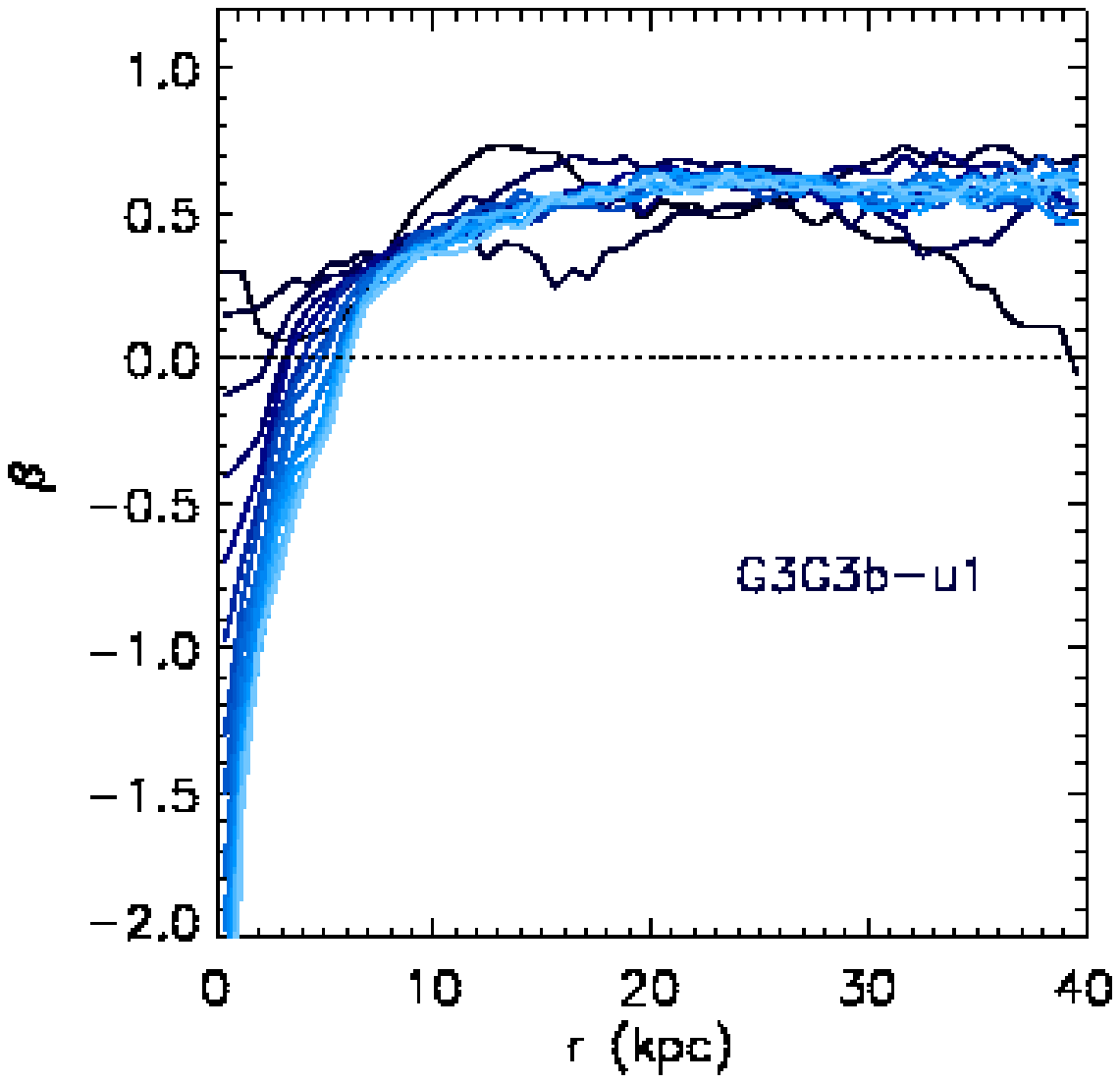}}
{\includegraphics{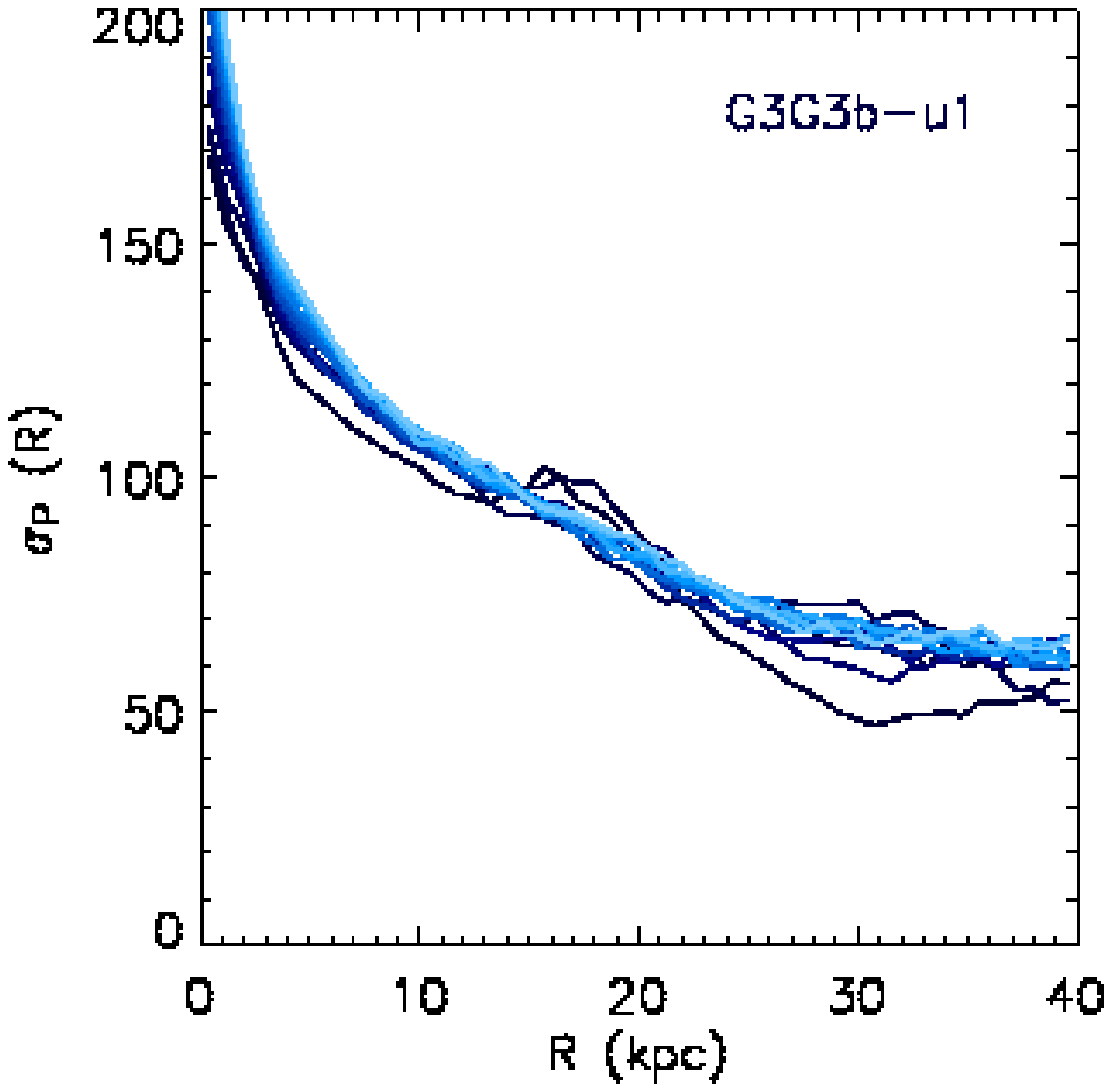}}
\vskip 0.2cm
\caption{Long-term stability.
The evolution of the $\beta$ and $\sp$ profiles in time after
the merger ($t=2.5$ Gyr) till $t=6$ Gyr. The curves are plotted at intervals
of 0.25 Gyr; they start black (dark blue) and become light blue progressively
in time. The profiles are stable at least until 3.5 Gyr after the merger.
}
\label{fig:time}
\end{figure}

% validity of the Jeans equation}
One can evaluate to what extent the simulated merger remnants obey the Jeans
equation, given their deviations from spherical symmetry, variations in time, 
and possible other deviations from full equilibrium.
Given in each remnant our measured values for 
$\alpha(r)$, $\beta(r)$, $\gamma(r)$ and $\sigma_r(r)$,
we have used the spherical Jeans equation to estimate the total mass 
$M_{\rm Jeans}(r)$ within radius $r$, in comparison with the true $M(r)$.
We find near $3\Re$ for the 10 remnants at about 1 Gyr after the merger
$M_{\rm Jeans}(r)/M(r) \simeq 0.9 \pm 0.15$. This is one possible measure
of the degree of validity of the spherical Jeans equation in these
systems.

%-------------------------------------------
%6
\section{Parametrizing the velocity-dispersion profiles}

One way to quantify the declining velocity dispersion profile is by
fitting a power-law with slope $\gamma$ to $\sp^2(\rp)$ in a given 
radius range.
The slope has been determined for each of the 60 individual profiles
by linear regression on equally-log-spaced
interpolation points in the range 0.5-$6\Re$.
The mean and standard deviation are 
$\gamma = 0.53 \pm 0.16$ for ``all" the stars,
and $\gamma = 0.61 \pm 0.22$ for the ``young" stars.
The PN data for the 4 observed galaxies, in the same radius range, give
$\gamma_p = 0.59 \pm 0.08 \pm 0.10$, for the mean, error on the mean 
(using error propagation), and scatter.  
There is an agreement between the simulations and the data, both for
``all" the stars and for the ``young" stars.

%The same analysis but in the range 1-$6\Re$ gives
%For ``all" the stars $\gamma_p = 0.60 \pm 0.21$.
%For the ``young" stars $\gamma_p = 0.70 \pm 0.27$.
%The PN data for the 4 observed galaxies, in the same radius range, give
%$\gamma_p = 0.84 \pm 0.22 \pm 0.23$, for the mean, error on the mean 
%(using error propagation), and scatter.

The $\sp$ profiles of the four observed galaxies were put together in
Fig.~2 of the paper for illustrative purposes. The scaling procedure,
explained in Methods, is robust to the exact method of scaling
and the range of data used for it.
We tried different minimum radii for the stellar
data used compared to the fiducial $0.2\Re$.
The resulting multiplication factors to get $\sp=1$ at $\Re$ 
for the 4 galaxies (NGC 4494, 3379, 821, 4697 respectively) are as follows.
For $r>0.2\Re$: (0.00779, 0.00761, 0.00637, 0.00714),
for $r>0.6\Re$: (0.00758, 0.00740, 0.00656, 0.00708),
for $r>1.0\Re$: (0.00788, 0.00806, 0.00623, 0.00685).
Using the PNs only we obtain:
(0.00804, 0.00779, 0.00656, 0.00681).
The relative changes are of a few percent only.

%--------------------------------
%7
\section{The full LOSVD and the $h_4$ moment}

We find that our sample of merger remnants spans a range of LOSVD's
similar to the observed sample, as traced both by stars and by PNs. 
This is illustrated first in \fig{losvd} which shows the LOSVD of three
of our merger remnants in comparison with three observed galaxies,
including negative and positive kurtosis (or $h_4$).
  
\begin{figure}[t] %7   Gary 
\vskip 5.5cm
%{\special{psfile="veldis.sbc_3379.ps" hscale=28 vscale=28 
%hoffset=-20 voffset=-50}}
%{\special{psfile="vdis_G3r400z_4697.ps" hscale=28 vscale=28
%hoffset=140 voffset=-50}}
%{\special{psfile="vdis_Sbc_o3_90z_5128.ps" hscale=28 vscale=28
%hoffset=300 voffset=-50}}
{\includegraphics{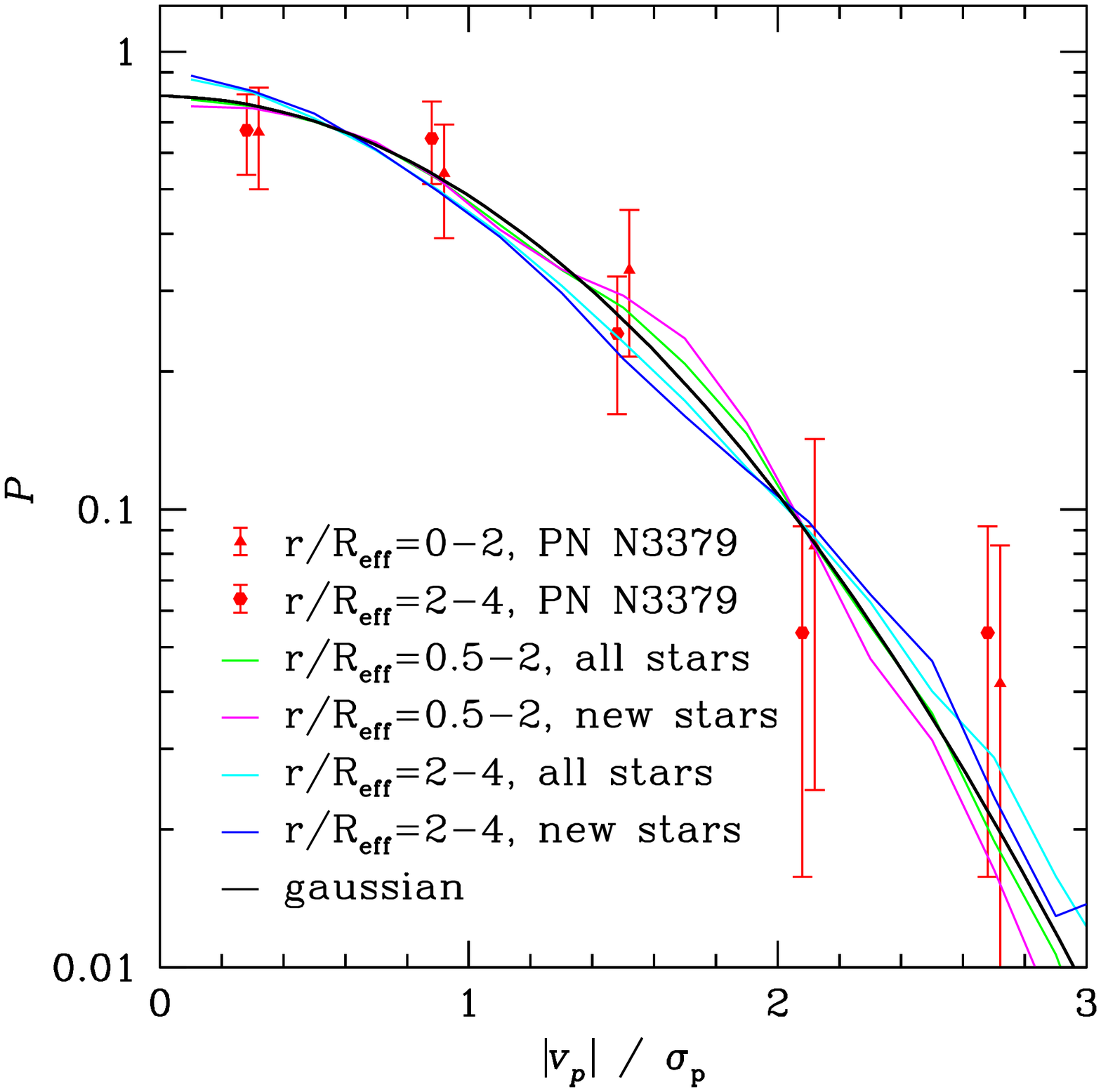}}
{\includegraphics{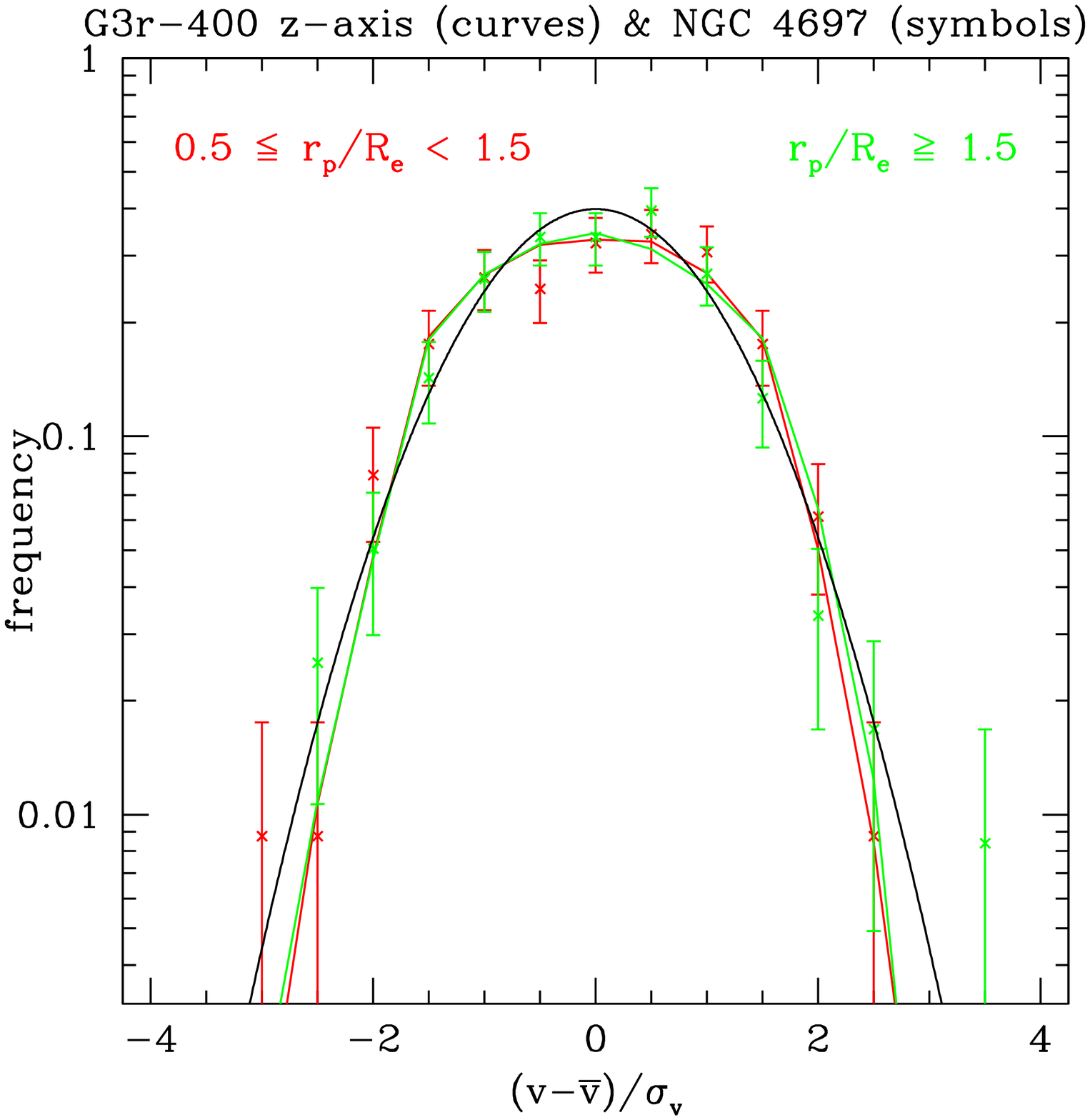}}
{\includegraphics{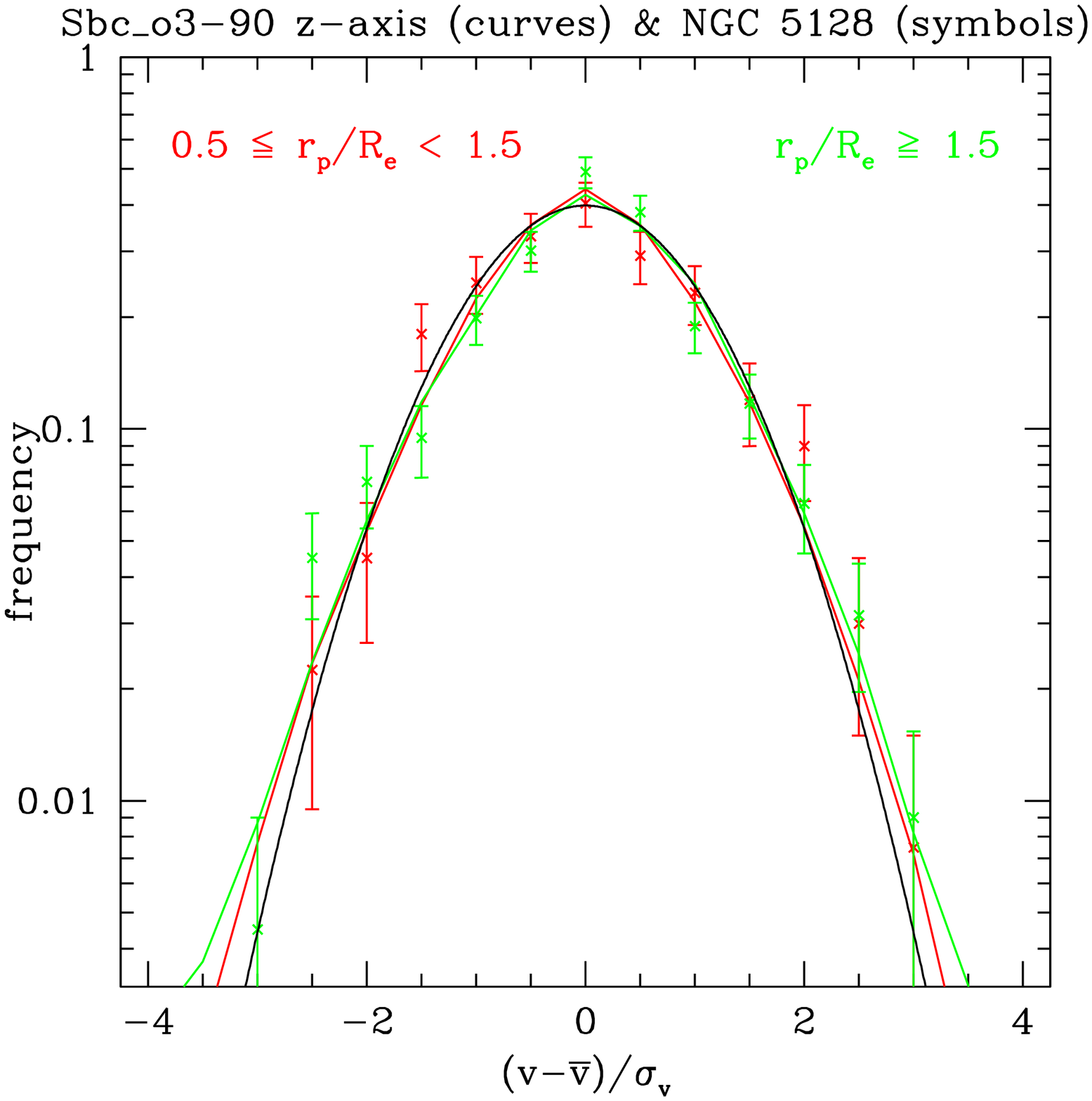}}
\vskip 0.2cm
\caption{Line-of-sight velocity distribution $P(|\vp|/\sp)$ (LOSVD)
of simulation versus data.
The black curve marks a Gaussian distribution.
The LOSVD is computed in two radial bins, as marked, within which the variation
of $\sp$ is limited.
The data is for N3379, N4697 and N5128 from left to right respectively.
The deviations from Gaussian are small, with a tendency for negative
and positive kurtosis in N4697 and N5128 respectively.
The curves are LOSVD from three simulations.
Left: the fiducial Sbc merger, stacked views from three orthogonal directions
at the final time, with a near Gaussian LOSVD.
Middle: a face-on view of a retrograde merger remnant (G3r) 
showing a negative kurtosis.
Right: a face on view of another Sbc merger remnant showing a positive
kurtosis.
We learn that a high anisotropy, $\beta \sim 0.5$ beyond $\Re$,
does not necessarily imply strong deviations from a Gaussian distribution.
}
\label{fig:losvd}
\end{figure}

We computed the central values of the fourth moment, $h_4$,
for the stars in the simulated remnants
using narrow radial bins, trying to mimic the typical observer's procedure. 
We made sure that the result within $\Re$ is insensitive to the
actual bin size (differences less than 10\%) once the bins are significantly
narrower than $\Re$.
\Fig{h4} shows the profiles of $h_4$ for all the simulated remnants stacked
together. The most relevant curve refers to the average over
many random lines of sights. The central value is $h_4 \simeq 0.03 \pm 0.05$.
The average within $\Re$ is $h_4 \simeq 0.02 \pm 0.04$.
In the outer range relevant to PNs, $1-4\Re$, it is $h_4 \simeq 0.01 \pm 0.05$.
However, more extreme values can be obtained if the line of sight
happens to be along one of the principal axes.
Along the major axis the central $h_4$ could be $\sim 0.1$ or larger,
and along the minor axis it could be $\sim -0.7$ at some radii and even
more negative.

Comparing the gas-poor G mergers ($\sim 20\%$ gas)
and the gas-rich Sbc mergers ($\sim 50\%$
gas) we find similar results for the average $h_4$ along random lines of
sight, but the G cases tend to show less extreme values when viewed along
the major or minor axes, reflecting less important central disk components
in these cases, as expected.

\begin{figure}[t] %8
\vskip 7.6cm
%{\special{psfile="stacked-h4-all-both.eps" hscale=80 vscale=100 
%hoffset=35 voffset=20}}
%{\special{psfile="n3379_h4.ps" hscale=33 vscale=33 
%hoffset=270 voffset=-48}}
{\includegraphics{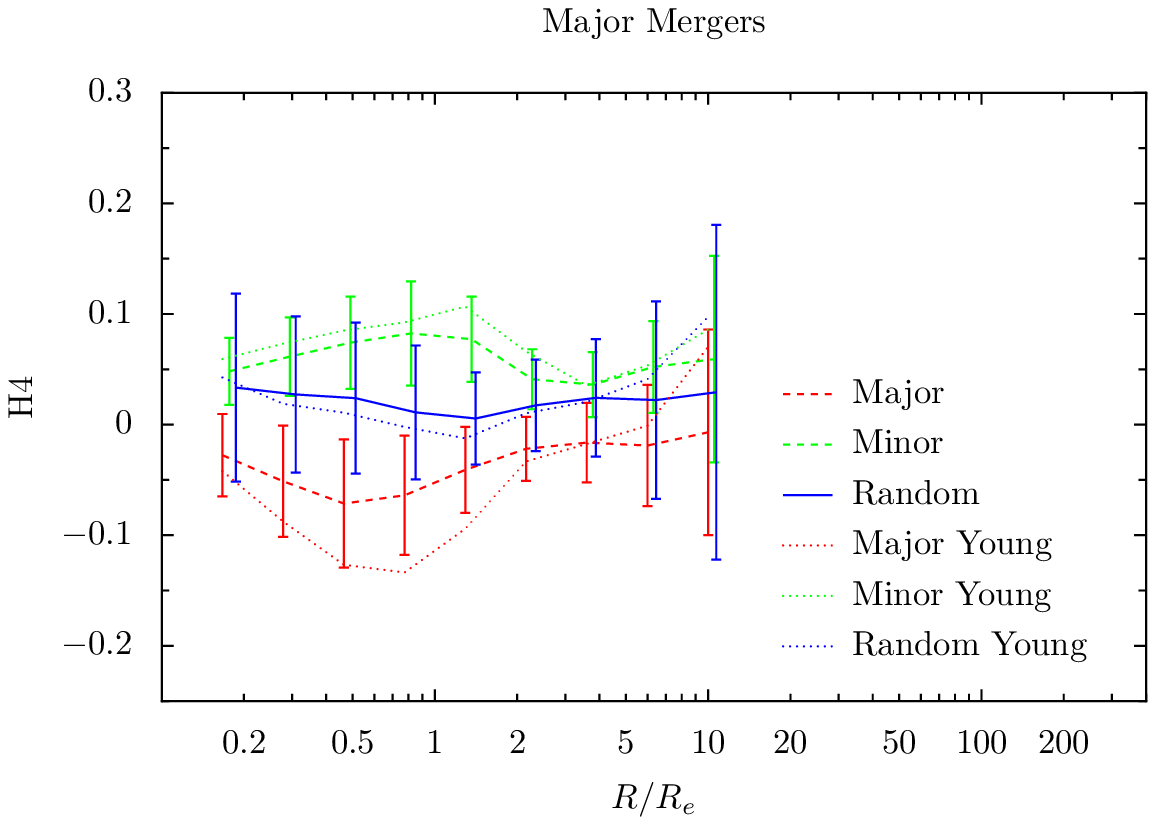}}
{\includegraphics{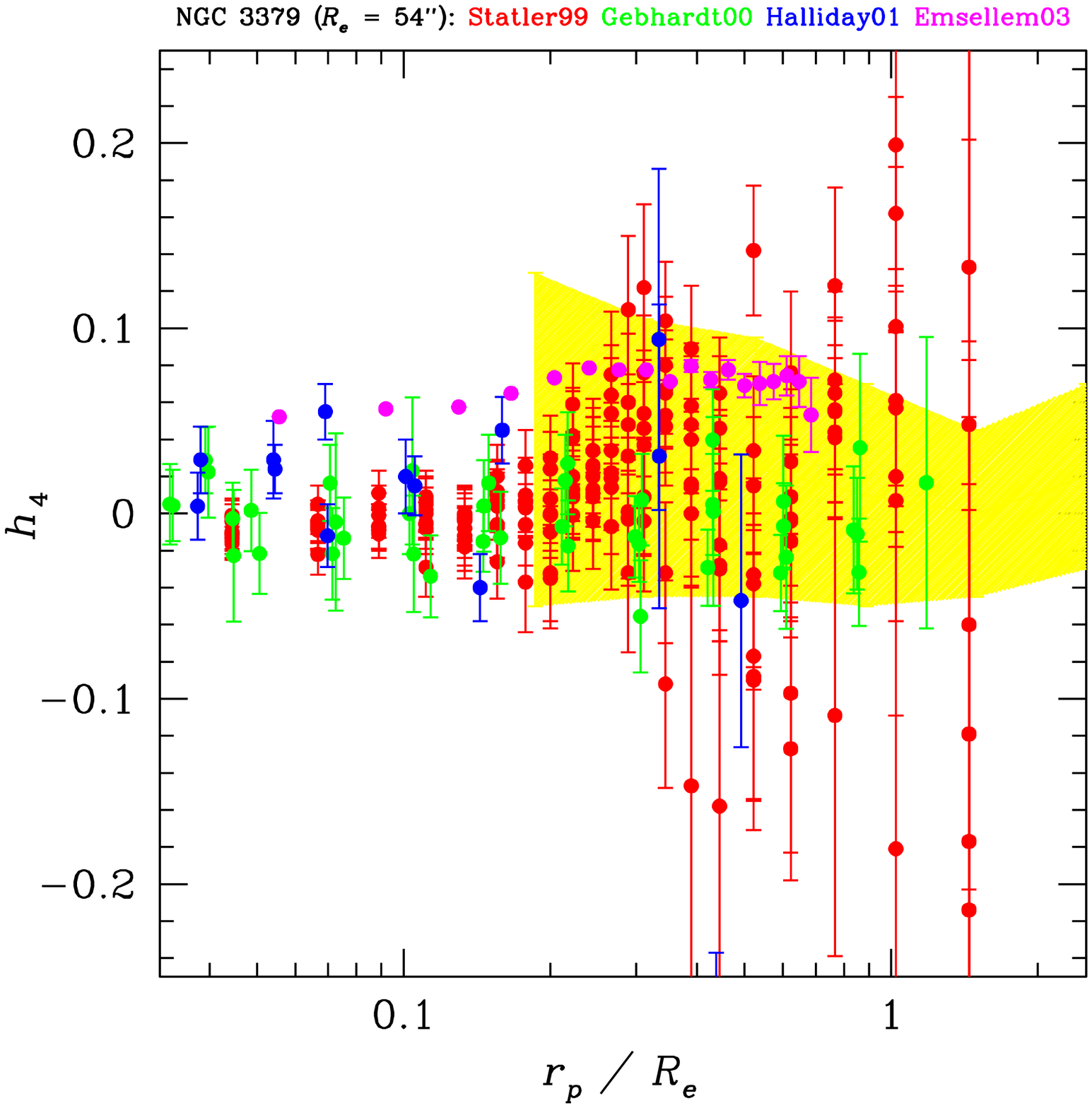}}
\vskip 0.2cm
\caption{The 4th moment of the LOSVD.
Left: Profiles of $h_4$ for all simulated remnants stacked together,
``all" and ``young" stars.
The line of sight is either random (blue), or along one of the principal
axes of the inertia tensor (as labeled). The error bars mark $1\sigma$ scatter.
Right: Profiles of $h_4$ for NGC 3379 from 4 different sources.
The $1\sigma$ range for the ``random" view of the simulated remnants   
is shaded yellow.
There is good agreement between the simulations and the observed galaxies.
}
\label{fig:h4}
\end{figure}
%stacked-h4-major-g.eps
%%stacked-h4-major-sbc.eps

These results are consistent with those of other simulations of "dry"
mergers\cite{bendo00,cretton01}, $h_4 = 0.02 \pm 0.04$.

The simulated $h_4$ is consistent with the observed $h_4$ for the
integrated stellar light in ellipticals.
This includes the typical values\cite{bender94}
$h_4 \sim 0.00 \pm 0.04$ 
and those for the four galaxies of the current 
Letter\cite{bender94,statler99,pinkney03}.

\Fig{h4} also shows the $h_4$ profile for NGC 3379 as measured by four
different observations. Within $0.3 \Re$ they span the range
$h_4 \simeq 0.0 \pm 0.05$.
The SAURON results\cite{emsellem04}
are slightly on the high side, ranging from $\sim 0.04$ to $\sim 0.08$.
At $r>0.3\Re$ the $h_4$ measurements are noisier; they extend to values
above +0.1 and below -0.1. 
In SAURON, $h_4$ is $0.080-0.085$ when averaged over circular bins,
with plenty of azimuthal scatter which may be due poor spectral resolution.
The SAURON $h_4$ profile of NGC 821 is similar,
with $h_4 \simeq 0.05$ within $\Re$.
The measured $h_4$ profiles in these two galaxies are thus consistent
with the distribution of $h_4$ profiles in the merger remnants.

Despite the general agreement with the observed values of $h_4$, we
stress that our sample of merger simulations is not supposed
to be a fair sample of {\it all\,} the mergers that lead to today's 
ellipticals. Clearly, not all ellipticals, maybe not even most of them,
had recent wet major mergers of disk systems.  However, we demonstrated
that the key properties relevant to the present Letter, such as the 
steep density profiles, high $\beta$ and low $\sp$ at {\it large\,} radii, 
are robust to wet and dry mergers, major and minor.
The simulations do not need to be as representative in terms
of the inner structure, where the wet mergers tend to lead to a density peak
and a small disk, as long as these inner features do not affect the key 
features at large radii.  In turn, we do not use here the
simulated properties in the {\it inner\,} regions as a diagnostic for the
physics involved in the formation of elliptical galaxies, being dry or wet,
etc.

One lesson from our simulations is
that a high anisotropy, $\beta \sim 0.5$ beyond $\Re$,
does not necessarily imply strong deviations from a Gaussian distribution.
This 
suggests %D3 
%D3 confirms the suspicion 
that the assumed\cite{gerhard93}
tight correlation between $h_4$ and
$\beta$ is model dependent and uncertain in several ways:
(a) it is based on the assumptions of spherical symmetry and 
a specific density profile,
(b) it is based on uncertain and limited orbit-library modeling, 
(c) it assumes that inner stars and outer PNs are the same population,
and (d) $h_4$ is also related to rotation.
We therefore do not draw conclusions from the
predictions of $\beta$ based on $h_4$.

%--------------------------------------
%8
\section{Matching other properties of ellipticals}

\no{\bf Fundamental Plane}.
The merger remnants qualitatively agree with the Fundamental Plane of
elliptical galaxies. This is demonstrated in \fig{fp}. The deviations
are typically at the $1\sigma$ level.
This is work in preparation. 

\begin{figure}[t] %9 Matt
\vskip 12cm
%{\special{psfile="fp_all_v3.eps" hscale=50 vscale=50 hoffset=60 voffset=-30}}
{\includegraphics{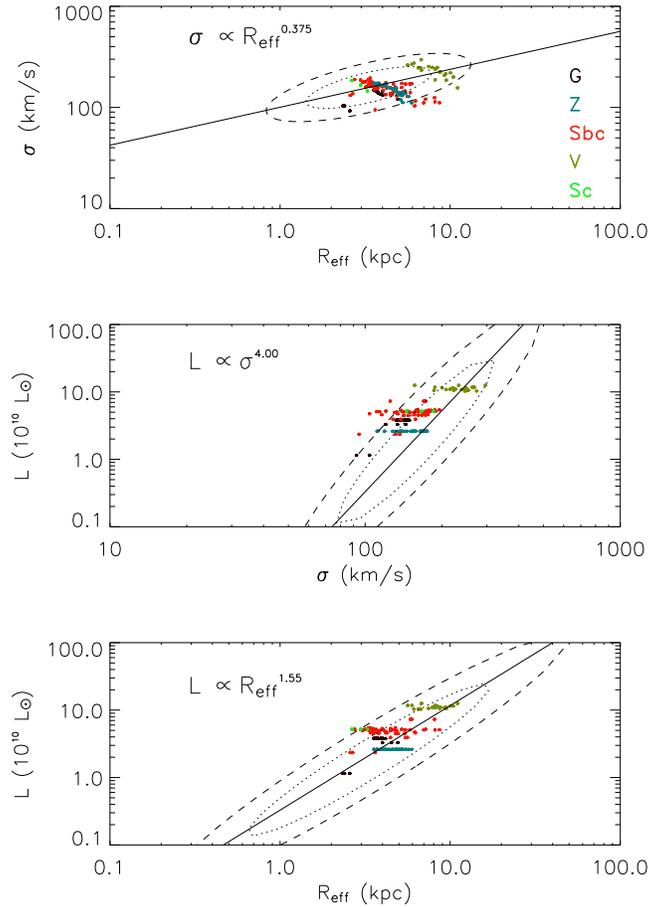}}
\vskip 0.2cm
\caption{The global structure properties of the simulated merger remnants 
(symbols, colored by the merger type) 
in comparison with the Fundamental Plane distribution of elliptical
galaxies in SDSS (1$\sigma$ and 2$\sigma$ contours).
The dispersion velocity $\sigma$ is measured at the central regions.
The luminosity is derived from the stellar mass assuming an effective $M/L=3$.
}
\label{fig:fp}
\end{figure}

\mno{\bf Surface-brightness profile}.
The simulated stellar surface-brightness profiles resemble the 
de Vaucouleurs profile remarkably well from below $\Re$ to several 
$\Re$ (Top panel of Fig.~2 of the Letter).
Some of our simulations produce a central stellar excess ($<1$kpc)
due to gas settling in an inner disk at the remnant center (which might be
prevented once AGNs are included). This explains the negative values of
$\beta$ at small radii in 
many %D3
%D3 some 
of the simulations. Such an excess is
not seen in typical ellipticals, but is seen in observed
mergers\cite{rothberg04,veilleux02}.
This inner excess does not
seem to affect the kinematics of the relevant, loosely bound stars 
beyond $\Re$.
The line-of-sight velocity dispersions are not
directly affected by the properties of the material at smaller radii.
Indeed, we find no correlation between the inner excess and the
anisotropy of the halo orbits.
The small excess is more pronounced
in the wet Sbc mergers than in the dryer G mergers, and more
so for the ``young" stars than the ``old" ones.
The fact that the low $\sp$ and steep $\gamma$ are also seen for ``all" the
stars and for the G mergers indicates that the central excess has no
important effect on our results.

\mno{\bf Ellipticity}.
The remnants of the 1:1 mergers provide projected axial ratios near $\Re$
that typically range from 1:1 to 1:2, not unlike observed ellipticals.
In 3D, the median configuration is 1:0.95:0.5, namely oblate,
but there are triaxial and prolate cases as well, in the ballpark
of
most %D3
observational findings\cite{dezeeuw91,ryden92,alam02,statler03}.
Retrograde mergers are naturally somewhat rounder than the more aligned
prograde mergers.
  
\mno{\bf Rotation}.
The simulated remnants may have a rotating gas-dominated disk at their centers,
but they are typically far from rotational support at $\Re$ and beyond.
For example, our fiducial Sbc case has near $\Re$ 
an axial ratio $c/a \simeq 0.5$ (or $\epsilon \simeq 0.5$).
The edge-on view gives a maximum of $V/\sigma_0 \simeq 0.25$ at $1.5\Re$,
well within the range of the estimates for the 4 observed ellipticals:
$V/\sigma_0 = 0.15-0.29$ (A. Romanowsky, private communication).
An isotropic case would have given  
$V/\sigma_0 = [\epsilon/(1-\epsilon)]^{1/2} \sim 1$.
Then $(V/\sigma_0)^* = (V/\sigma_0)/(V/\sigma_{0,{\rm isotropic}}) \sim 0.25$,
namely, the system ellipticity is dominated by the triaxiality of 
the velocity dispersion tensor, not by rotation (consistent with 
non-dissipative merger simulations\cite{naab03}).  
%Similar results are valid for the G mergers, 
%with a typical maximum $V/\sigma_0 \simeq 0.3$ and 
%$(V/\sigma_0)^* \simeq 0.5$.

% Compared to the local $\sigma$ at the same radius, it is V/sigma ~0.3
% %(growing to ~0.6 at 10Re)
% %Still, the sigma_v viewed face-on at 1.5Re is a factor ~1.8
% %smaller than that viewed adge-on!  A large scatter due to the
% %triaxiality dependening on the direction of view. This is the main
% %cause of scatter in our simulated sigma_v results, Fig. 2.
% %Romanowsky:
% % N821:  (v/sigma_0) = 0.24, e = 0.24  ->  (v/sigma_0)* = 0.4   a4 = 2.5
% % N3379:               0.29      0.14                     0.7        0.2
% % N4494:               0.15      0.15                     0.4        0.3
% % N4697:               0.23      0.3                      0.4        1.4

In another merger simulation\cite{springel05}
the remnant looks like a spiral almost
immediately after the merger.  This is perhaps due to their somewhat
pathological orbit, where the spins and angular momenta are aligned.
It is true that our Sbc merger remnants have certain disk-like
qualities because they started out with a high gas fraction and an extended
gas disk.  However, the G remnants have less disks in them,
implying that the high $\beta$ and low $\sp$ are not
a result of the diskiness of the remnant.

%D3
Yet another set of merger simulations\cite{gonzalez05} indicates a tendency
for more tangential orbits at large radii. This is likely to be due to their
wide angle and slow orbits, in which the tidal effects at pericenter are
weaker than in our more radial and faster collisions. 

\mno{\bf Dark-matter density profile}.
The simulated DM profiles in the relevant range between $\Re$ and a few $\Re$
are typically of log slope slightly flatter than -2. This is in the
ballpark of what might be expected from CDM haloes below $\sim 0.1 \Rv$
subject to the baryonic dissipative 
effects\cite{gnedin04}, %D3
as mentioned in the text.
(This could be compared to the unperturbed NFW profile of haloes in DM-only
simulations, which flattens to $\sim -1.5$ near $\Re$.)
%D3 xxxxx check 

%%%%%%%%%%%%%%%%%%%%%%%%%%%%
%9
\section{Scatter induced by triaxiality}

\Fig{snaps_sbc} illustrates the triaxial shape of our typical merger
remnant. \Fig{triaxial} demonstrates the effects of triaxiality on the
velocity dispersion profile.

\begin{figure}[t] %10
\vskip 7.0cm
%{\special{psfile="sigma2d.fiducial.eps" 
%hscale=50 vscale=50 hoffset=75 voffset=-15}}
{\includegraphics{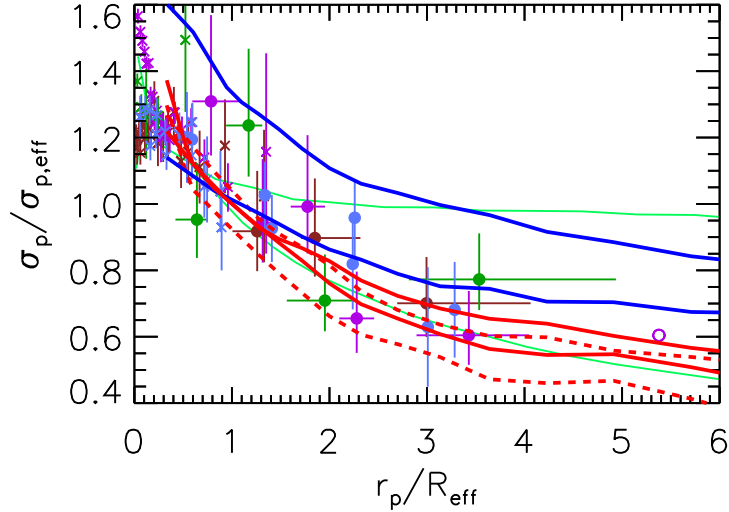}}
\vskip 0.2cm
\caption{Effect of triaxiality and rotation.
Projected velocity dispersion in the Sbc merger remnant of 
\fig{snaps_sbc} as viewed ``face-on" (lower) and ``edge-on" (upper).
Notation as in Fig.~2 of the Letter.
The triaxiality in the velocity dispersion tensor leads to a $\sim 20\%$
difference between the two projection directions.
The $\sp$ profiles from both views of ``all" the stars are consistent with the
observed $\sp$.  The edge-on view of the ``young" stars is also
in good agreement with the observed $\sp$, while the face-on $\sp$
generally lies below the data.
While rotation is part of the anisotropy of the tangential component,
the simulated system is not supported by rotation:
its edge-on rotation velocity is
only $\simeq 25\%$ of the circular velocity, with $V/\sigma_0 \simeq 0.2$
at $\gsim \Re$, and with $V/\sigma \simeq 0.3$ below $\Re$
rising to $V/\sigma \simeq 0.5$ beyond $2\Re$.
}
\label{fig:triaxial}
\end{figure}

While the four ellipticals addressed here have low, declining velocity
dispersion profiles, there are other ellipticals which show
higher or flatter dispersion profiles.
As mention in the text, such profiles may be the result of:
(a) a specific projection angle in a triaxial or rotating system,
(b) a high angular-momentum merger,
(c) a retrograde merger,
(d) a merger with less dissipation,
or (e) probing with tracers different from PNs, such as globular clusters
(of a flatter density profile and lower anisotropy).
The high $\sp$ cases may be associated with more massive galaxies
and with centers of groups --- a subject of future work.

%---------------------------
%10
\section{On the PN age}

One may be concerned about the validity of the assertion that the bright PNs
are likely to be younger than 3 Gyr based on the theoretical work of Marigo
et al.\cite{marigo04}.

We should stress first that the issue of whether the PNs are young or not is
{\it secondary} to our results, corresponding to only a small additional
reduction of less than 10\% in $\sp$. This is smaller than the uncertainties 
and variation among the galaxies, and smaller than the variations due
to the merger-orbit parameters or projection direction. Our key result, 
of radial anisotropy, is even more robust, as seen in Fig.~1 of the Letter
and as demonstrated above.  

Still, the theoretical analysis seems straightforward and
we are not aware of convincing theoretical arguments against its validity. 
The similar luminosity function of PNs in different ellipticals
is an empirical result which should rather be explained (e.g. by a 
similar merger history for these galaxies, including many minor mergers), 
but it does not seem to justify ignoring the theoretical predictions 
based on stellar theory. It is therefore fair to consider the PN young age as a
possibility.

We may comment in passing that the hypothesis
that the PNs are blue stragglers produced by stellar collisions,
may perhaps work for globular clusters, but it is hard
to see how the stellar collision rate could be high enough at the 
outskirts of elliptical galaxies. Perhaps a recent merger would help...
The idea of blue stragglers may work if they are produced by 
mass transfer in common binary systems\cite{ciardullo05}.

%-----------------------------
%11
\section{Rising versus constant $\beta$}

Is the gradual rise in $\beta(r)$ responsible
for a steeper decline in $\sp$? 
This seems consistent with the trend seen in the comparison of ``dry" and 
``wet" simulations in \fig{dry}, for which $\beta$ is a constant and 
rising respectively.

However, a test using the spherical Jeans equation indicates that
low $\beta$ values below $\Re$ have negligible effects on $\sp$ beyond $\Re$.
\fig{beta_rise} compares two solutions of the Jeans equation, one with 
$\beta=0.5$ at all radii, and the other with $\beta=0$ at small radii,
smoothly rising near $\Re$ to $\beta=0.5$ at large radii following an
assumed functional form\cite{mamon05a}.
For example, $\beta=0.25$ at $1.4\Re$.
The shape of the $\beta$ profile is similar to the average shape
in our simulated remnants.
After matching the solutions at $\Re$, the $\sp$ profiles are almost identical
at all larger radii. Based on this example, we suspect that the rising 
nature of the profile may not to be the key factor driving the decline 
in $\sp$.  

We recently learned from A. Romanowsky (private comm.) about the $\beta$ 
profile in their preferred orbit-library solution to N3379: 
it has a negative value of $\beta=-0.3$ at $\Re$, 
and it rises toward $\sim +0.5$ but only beyond $5\Re$. 
While being rising in part of the relevant range, 
it does not seem to reach high enough values there, and therefore
does not support a massive DM halo. 
In comparison,
in many of our simulations, $\beta$ is positive at $\Re$, it is
close to $0.5$ already at $2\Re$, and it keeps rising toward larger radii.

\begin{figure} %11  Gary: fixed beta vs rising beta
\vskip 5.5cm
%{\special{psfile="beta_rise.ps" hscale=40 vscale=40 hoffset=110 voffset=-110}}
{\includegraphics{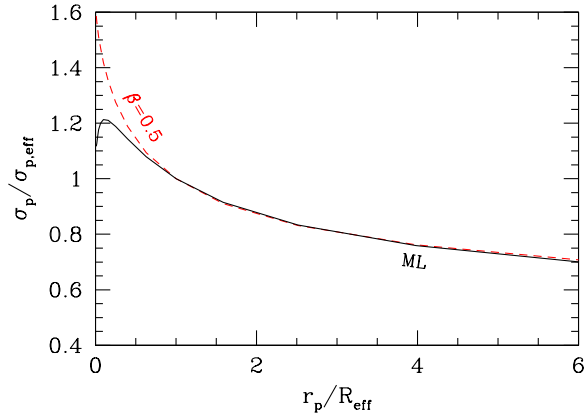}}
\vskip 0.2cm
\caption{Rising versus constant $\beta$.
A comparison of solutions to the spherical Jeans equation;
one with $\beta=0.5$ at all radii and the other with a rising $\beta(r)$
from zero at small radii to $0.5$ at large radii, with the transition
near $\Re$.
}
\label{fig:beta_rise}
\end{figure}

%---------------------------
%12
\section{Why was the DM solution missed before?}

Using merger simulations, we provide ``a" fully specified solution,
consistent with what we know about DM haloes in other galaxies and in the
$\Lambda$CDM scenario, which seems not to violate all the relevant 
observed aspects
within the given uncertainties.  Independent of whether this solution is 
indeed ``the" solution for elliptical galaxies, it is
claimed\cite{romanowsky03} (R03) to have been missed in the original
application of the orbit-library maximum-likelihood algorithm.

We cannot provide a definite
answer to why this solution was missed; the published material 
(as well as private communication with the authors) provided only partial
information for evaluating the method or for fully characterizing their
preferred solutions.  
A sample output from our simulations can provide a useful testbed for 
their method.

However, it is not at all surprising that a complex maximum-likelihood
method based on an orbit library may miss some of the solutions. 
We can only raise several ideas for where such method could go wrong,
as follows: 
\begin{enumerate}
\item
Assuming a stationary potential. The outer system may still be evolving (see
\fig{radial_origin}). We find in our Sbc simulations that the Jeans equation
is typically valid to $\sim 10\%$ accuracy, in the sense that the total 
mass within $1-3\Re$ as derived by the Jeans equation is $\sim 90\%$ of
the true value. In a few cases the deviation can be larger.
\item
Assuming spherical symmetry. Ellipticals can appear circular in one projection
while actually having an axial ratio of 1:2 (see \fig{snaps_sbc}).
\item 
Assuming density profiles with a specific, limited shape, such as NFW for
the DM.  
There is no reason to believe that the DM haloes in ellipticals,
under the influence of dissipation within them, still obey their
dissipationless NFW shape near $\Re$\cite{gnedin04}.
In fact, our simulations demonstrate that a major merger of NFW haloes
including disks may lead to slight deviations from NFW 
in the relevant range (see Fig.~1 of the Letter, 
showing a power-law of slope $\sim -2$ in the relevant range).
\item 
Their orbit-library method heavily depends on constraints from the values
of higher
velocity moments, such as $h_4$, as determined from the central stellar diffuse
light. These constraints may be misleading if the PNs detected at
large radii actually represent a population that is kinematically different
from the central stars. 
\item
The problem may lie in some incompleteness of the orbit library used,
or in the way by which it is composed for best fit. There were claims
in the literature about failure to converge when the number of 
orbits is increased\cite{valluri04,cretton04,richstone04}.
One may suspect that the R03 method is subject to such problems.
\item
A multi-parameter (or ``non-parametric") maximum-likelihood method may
easily miss some of the solutions.
%\item
%In their Jeans modeling, R03 assume a constant $\beta$, while our models
%typically have a rising $\beta$ profile.
\end{enumerate}

%---------------------------------------
%13
\section{Robustness and uniqueness of the results}
  
Unfortunately, full exact predictions of the $\Lambda$CDM scenario are
beyond what can be achieved by current methods. We therefore use
state-of-the-art simulations to study in detail a sample of merger
cases. This sample, based on plausible assumptions,
provides DM-dominated systems which are
consistent with the observed ellipticals in {\it all\,} relevant testable
respects, and it reproduces the puzzling
low velocity dispersions and their variations among different
galaxies.  This by itself is an important step forward, given that previous
work failed to find such a solution and thus challenged
the whole cosmological scenario.  Demonstrating that the low velocities
are {\it not necessarily fatal\,} for $\Lambda$CDM is a step
forward by itself.

However, the variety of cases
simulated provide reasonably compelling evidence for the robustness of
the key phenomenon --- radial orbits --- indicating that this is indeed a
{\it generic\,} outcome of a large class of mergers in CDM-like scenarios.
In particular, we have demonstrated robustness to {\it dissipation},
{\it mass ratio}, the presence of a {\it bulge\,} 
and several other merger characteristics, and have gained insight into the 
{\it generic origin\,} of the effect via tidal processes.

An alternative solution has been proposed\cite{napolitano05},
appealing to a low halo concentration parameter $C$.
Using our Jeans-equation algorithm, we find
that even a $C$ as small as 30\% of the mean $\Lambda$CDM value
fails to fit the $\sp$ profile of NGC 3379 unless 
$\beta \sim 0.5$ near $\Re$ and
beyond.  The fit proposed\cite{napolitano05} 
requires an unrealistic $C$, five times
smaller than the $\Lambda$CDM value.  
This indicates that our solution, based on
radial stellar orbits, is the only viable solution proposed so far in the
literature within the $\Lambda$CDM cosmology.  

We have demonstrated 
%D more conclusively 
that the solutions
emerging from our sample merger simulations are representative of
a large class of mergers likely to occur in a $\Lambda$CDM cosmology.
We conclude that the radial
anisotropies necessary for compatibility with the low $\sp$ observed 
are not really ``pathological" but generally expected. Then, 
triaxiality of the galaxies could play a major role
in observing low dispersions; recent star formation could in rare cases
also contribute."

%%%%%%%%%%%%%%%%%%%%%%%%%%%%%%%%%%%%%%%

%\addtolength{\baselineskip}{-0.05\baselineskip}
%\addtolength{\parskip}{-0.2\baselineskip}

%\bibliographystyle{natureedo}
%\bibliography{pn_astroph}

%%%%%%%%%%%%%%%%%%%%%%%%%%%%%%%%%%%%%%%
\end{document}